\documentclass[preprint,12pt]{aastex}
\usepackage{natbib}
\newcommand{\kms}{km s$^{-1}$}
\newcommand{\vsys}{systematic velocity}
\newcommand{\pa}{position angle}
\newcommand{\inc}{inclination}
\newcommand{\vrad}{$v_R$}
\newcommand{\vtan}{$v_\theta$}

\newcommand{\halpha}{H$\alpha$}
\shorttitle{Spiral Streaming in M51}
\shortauthors{Shetty et al}

\begin{document}
\title{Kinematics of Spiral Arm Streaming in M51}
\author{Rahul Shetty, Stuart N. Vogel, Eve C. Ostriker, and Peter
  J. Teuben}
\affil{Department of Astronomy, University of Maryland, College Park,
  MD 20742-2421} 
\email{shetty@astro.umd.edu, vogel@astro.umd.edu,
  ostriker@astro.umd.edu, teuben@astro.umd.edu}

\begin{abstract}

  We use CO and \halpha\ velocity fields to study the gas kinematics
  in the spiral arms and interarms of M51 (NGC 5194), and fit the 2D
  velocity field to estimate the radial and tangential velocity
  components as a function of spiral phase (arm distance).  We find
  large radial and tangential streaming velocities, which are
  qualitatively consistent with the predictions of density wave theory
  and support the existence of shocks.  The streaming motions are
  complex, varying significantly across the galaxy as well as along
  and between arms.  Aberrations in the velocity field indicate that
  the disk is not coplanar, perhaps as far in as 20\arcsec\ (800 pc)
  from the center.  Velocity profile fits from CO and \halpha\ are
  typically similar, suggesting that most of the \halpha\ emission
  originates from regions of recent star formation.  We also explore
  vortensity and mass conservation conditions.  Vortensity
  conservation, which does not require a steady state, is empirically
  verified.  The velocity and density profiles show large and varying
  mass fluxes, which are inconsistent with a steady flow for a single
  dominant global spiral mode.  We thus conclude that the spiral arms
  cannot be in a quasi-steady state in any rotating frame, and/or that
  out of plane motions may be significant.

\end{abstract}

\keywords{galaxies: individual (M51) -- galaxies: spiral structure --
  galaxies: kinematics and dynamics}

\section{Introduction}
Spiral arms are the dominant morphological features of most disk
galaxies.  From a theoretical perspective, two frameworks have been
proposed to describe the nature of the spiral arms: one is that the
spiral arms are generally long-lasting, or slowly evolving, and the
other is that the arms are transient features
\citep[e.g.][]{ToomreToomre72}.  Observational studies have yet to
show definitively whether the arms are evolving or long lived, though
it has been over 40 years since the landmark paper by \citet{LinShu64}
suggesting that spiral structure in galaxies is a long lived
phenomenon --- the Quasi-Stationary Spiral Structure (QSSS) hypothesis
\citep{Lindblad63}.  In the QSSS depiction, though material passes in
and out of the arms, the slowly evolving global pattern rotates with a
single angular speed that results from the excitation of global modes.
The spiral arms are formed from self-excited and self-regulated
standing density waves \citep{Bertin89a, Bertin89b, BertinLin96}.

However, interaction between a disk galaxy and a companion is another
explanation for the presence of spiral arms.  In such a framework, the
arms are transient features that are generated by the tidal
interaction \citep[e.g.,][]{ToomreToomre72}.  Any spiral arms existing
before the encounter are overwhelmed by the tidal driving
\citep{SaloLaur00I}.

Regardless of the origin of the stellar arms, gas in the disk will
respond strongly to the gravitational perturbations those arms impose.
Numerical studies have indicated that shocks can develop if the
relative speed between the spiral perturbation and the gas is large
\citep{Roberts69, ShuMilioneRoberts73, Woodward75}.  The presence of
dust lanes in the spiral arms and the enhancement in ionized emission
downstream, indicating regions of star formation, is attributed to this
shock scenario.  Such shocks are also thought to be the cause of the
well defined molecular arms seen in many grand design galaxies,
including M51.  Numerical and analytical studies have provided
predictions for the velocity and density profiles of the matter
affected by the spiral gravitational perturbation
\citep[e.g][]{LubowBalbusCowie86, KO02, GittinsClarke04}.

There have been numerous observational studies addressing the nature
of spiral structure that have focused on the gaseous components.
\citet{Visser80a} showed that steady state density wave models fit the
HI kinematics of M81 quite well.  \citet{Lowe94} used the modal theory
of density waves to describe the spiral pattern in M81.  Both
\citet{Rand93} and \citet{Aalto99} used observed molecular velocities
along 1D cuts on the major and minor axes of the grand design spiral
M51, and found qualitative agreement with the density wave models of
\citet{RobertsStew87}.  \citet{KunoNakai97} fitted observed CO
velocities from single dish observations to obtain gas streamlines.
The smooth shape of the velocity profiles led them to conclude that
galactic shocks do not exist in M51.  However, the study by
\citet{Aalto99}, using higher resolution interferometric data, found
steeper velocity gradients, supporting the presence of shocks.

Yet, other observational studies have suggested that the arms are not
long lived.  In fact, the classic kinematic study of M51, that of
\citet{Tully74sp}, found evidence for a transient pattern in the outer
disk, due to the interaction with its companion, but that a steady
state is probably appropriate for the inner arms.
\citet{ElmSeidenElm89} and \citet{Vogel93} also suggested the presence
of multiple pattern speeds.  \citet{HenryQG03} argued that the spiral
pattern may be a superposition of an $m$=2 mode and a weaker $m$=3
mode, suggesting a transient pattern for the arms of M51.

This paper presents a detailed study of the gaseous velocity structure
associated with the spiral pattern in M51.  In a future paper, we will
discuss and compare the spiral pattern in different tracers.  Here, we
use the CO and \halpha\ velocities to map the 2D velocity field in
M51.

Our study makes use of the full 2D velocity field in M51 from
interferometric CO and Fabry-Perot \halpha\ observations, rather than
just major and minor axis cuts.  Noting that variations in the
observed velocity field are mainly associated with the spiral arms, we
fit the observed velocity field to obtain the radial and tangential
components as a function of arm phase (i.e. distance perpendicular to
the arm).  We then analyze whether the fitted velocity field and
density maps are consistent with the predictions of steady state
theory.

In the next section we briefly describe our CO and \halpha\
observations.  In $\S$\ref{fitting} we describe the method we employ
to estimate the radial and tangential velocity components throughout
the disk.  Since our method is sensitive to the assumed values of the
\vsys, major axis \pa, and disk \inc\ with respect to the sky, in
$\S$\ref{methodtest} we present results from our effort to constrain
these parameters, and describe how errors could affect the fitting
results.  In $\S$\ref{results} we present and discuss the fitted
profiles of radial and tangential velocities for a range of radii.  We
then use the velocity and density profiles to empirically test
conservation of vortensity in $\S$\ref{vort_sec}.  Next, in
$\S$\ref{consmass}, we examine whether (quasi) steady state mass
conservation is applicable, as would be necessary for a QSSS
description.  Finally, in $\S$\ref{summary}, we summarize our
conclusions.

\section{Observations}

The CO and \halpha\ intensity and velocity maps are shown in Figures
\ref{COarm} and \ref{Haarm}, respectively.  The CO J=1-0 data for M51
were obtained in part from BIMA SONG (Survey of Nearby Galaxies).  The
observations and data reduction are described in \citet{Regan01} and
\citet{Helfer03}.  The SONG map is based on 26 pointings and has an
angular resolution of 5.8\arcsec\ $\times$ 5.1\arcsec.  Later, we
obtained data for 34 additional pointings, so that the spiral arms
were mapped as far as the companion galaxy to the north and to a
similar distance along the spiral arm to the south.  Additionally,
inner fields were mapped in a higher resolution array (B array),
yielding higher angular resolution.  The newer data were reduced using
the same procedures as described in \citet{Helfer03} for BIMA SONG.
Together the data sets cover 60 pointings.  The maps used for this
paper have variable resolution, reaching as high as 4\arcsec\ in the
inner spiral arms but degrading to 6\arcsec\ -- 13\arcsec\ in the
interarms and in the outer arms.

\halpha\ data were obtained with the Maryland-Caltech Palomar
Fabry-Perot, which covered the optical disk at an angular resolution
of 2\arcsec\ and a velocity resolution of 25 \kms.  The observations
and reduction are described in \citet{Gruendl96} and also
\citet{Vogel93}.  Both CO and \halpha\ intensity maps are obtained by
fitting Gaussian profiles to the spectrum at each location.  The
velocity maps indicate the velocity of the peak of the fit Gaussian
intensity.

Also shown in Figures \ref{COarm} and \ref{Haarm} are two lines
tracing logarithmic spirals.  The bright CO arm is well represented by
a logarithmic spiral with a pitch angle of 21.1\degr.  The weaker arm
also generally follows a logarithmic spiral, although, as will be
discussed, a number of its arm segments either lead or lag the
depicted line.  The logarithmic spirals will be discussed extensively
in the following sections.

\section{Estimation of Spiral Streaming Velocities \label {fitting}}

The observed line of sight velocity $V_{obs}$ can be decomposed as a
sum of terms involving the \vsys\ $V_{sys}$, the radial velocity
\vrad\, and the tangential velocity \vtan:
\begin{equation}
  {V_{obs}(R,\theta)=V_{sys}+[v_{R}(R,\theta)\sin(\theta-\theta_{MA})+v_{\theta}(R,\theta)\cos(\theta-\theta_{MA})]\sin i} ,
\label{vobs}
\end{equation}
where $R$ and $\theta$ are the galactocentric radius and azimuthal
angle, and $\theta_{MA}$ and $i$ are the \pa\ of the major axis and
\inc\ of the galaxy, respectively.  This equation does not include a
velocity component perpendicular to the disk.  The exclusion of the
vertical velocity component is reasonable since studies of face-on
grand design spirals indicate that the $z$-component of velocity is
less than 5 \kms \citep{KruitShostak82}, provided the disk has no
significant warp (we return to this issue in $\S$\ref{vsyssec}).

Inspection of the velocity maps indicates that the isovelocity
contours near the spiral arms tend to run parallel to the arms.  For a
disk in pure circular rotation and with a flat rotation curve, on the
other hand, the isovelocity contours of the projected velocity field
are purely radial.  It is evident that the velocity field of M51 is
significantly different from this sort of simple ``spider diagram,''
due to the non-axisymmetric perturbations associated with spiral
streaming.  Clearly, \vrad\ and \vtan\ vary with azimuth.

Previous estimates of streaming velocities have used observed
velocities near the major axis (where the projections of \vrad\
vanishes) to estimate \vtan, and velocities near the minor axis (where
the projections of \vtan\ vanishes) to estimate \vrad\
\citep[e.g.,][]{Rand93}.  However, much of the CO gas is organized
into GMCs and larger complexes known as GMAs \citep{Vogel88}.
Further, \citet{Aalto99} have found that the streaming velocities of
M51 GMAs in the same spiral arm have a significant dispersion.
Observations along a single cut, e.g. the major axis, sample discrete
GMCs and therefore may give a misleading estimate of the streaming
velocities.  As an alternative, an approach that fits a streaming
profile to all the observed velocities in an annulus as a function of
distance from the arm peak may better characterize the streaming
velocities.  Also, the gas surface density distribution varies
significantly at any distance from the peak (i.e. the gas is clumpy),
and so averaging parallel to the arm may better characterize the
variation in gas surface density as a function of arm distance.

\begin{figure}
\epsscale{1.7}
\includegraphics*[scale=0.95]{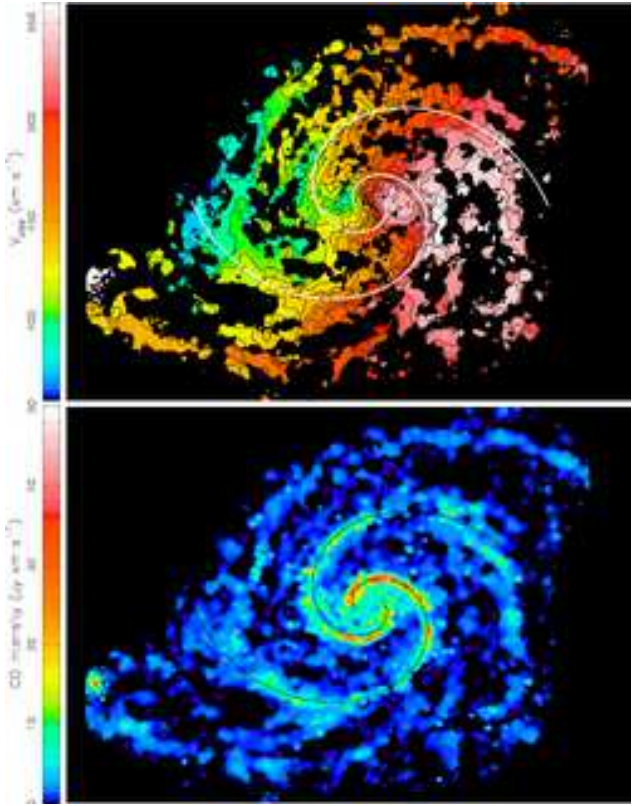}
\caption{\footnotesize CO (1-0) velocity-integrated intensity (bottom)
  and velocity (top) maps of M51.  Velocity contours increment by 10
  \kms, between 360 and 560 \kms. Overlaid lines are logarithmic
  spirals with a pitch angle of 21.1\degr, separated by 180\degr.}
\label{COarm}
\end{figure}

\begin{figure}
\epsscale{1.7}
\includegraphics*[scale=0.95]{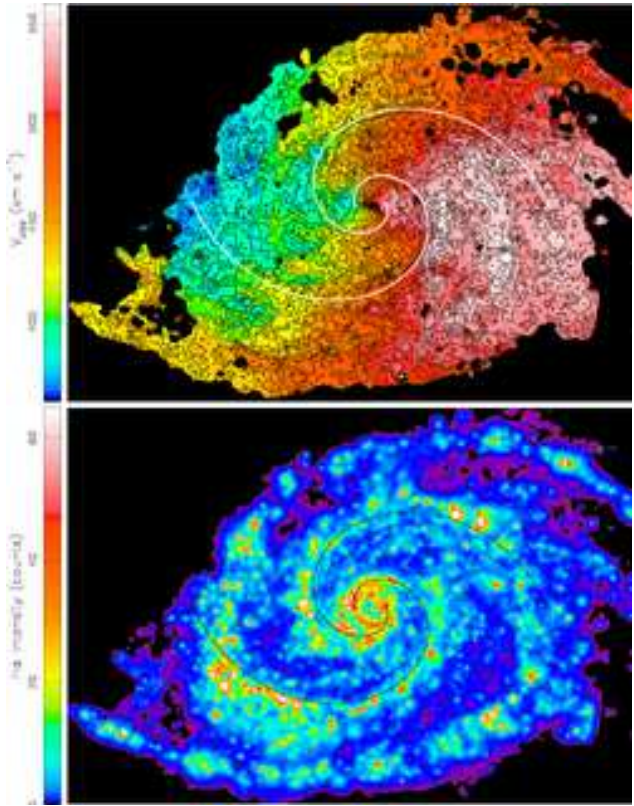}
\caption{\footnotesize \halpha\ velocity-integrated intensity (bottom)
  and velocity (top) map of M51.  The overlaid spirals, as well as the
  velocity contours, are as described in Figure \ref{COarm}.}
\label{Haarm}
\end{figure}

\begin{figure}
\epsscale{0.9}
\plotone{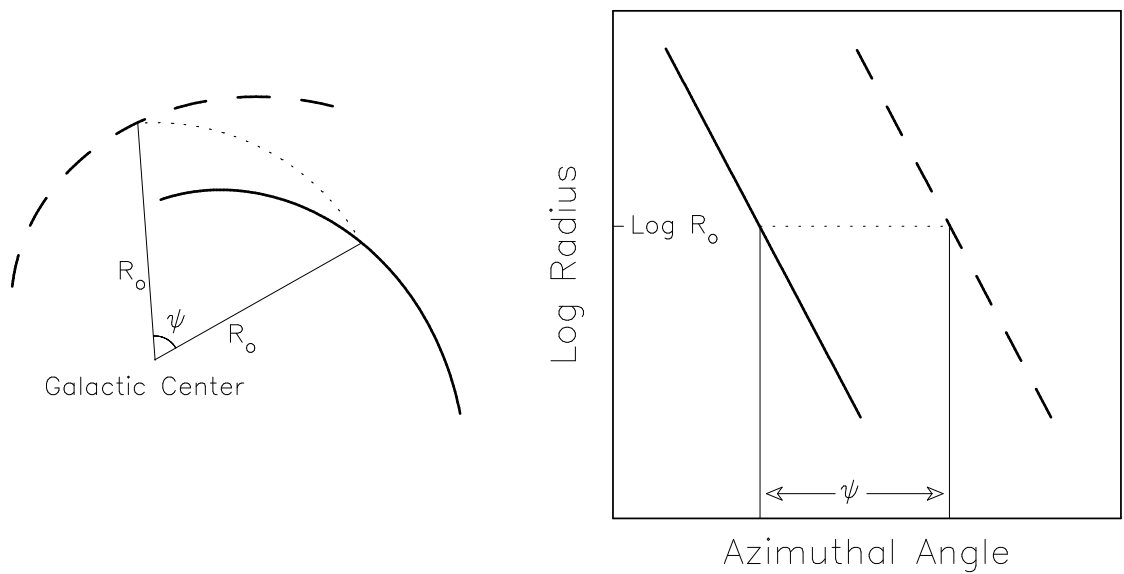}
\caption{\footnotesize Geometry depicting the spiral arm phase $\psi$.
  The diagram on the left is the geometry in the plane of the galaxy.
  The phase, $\psi$, represents the angular displacement between two
  locations with equal galactocentric radius $R_o$ for two congruent
  spiral segments.  The diagram on the right is the logarithmic polar
  projection of the geometry on the left, showing the corresponding
  spiral segments.}
\label{armphase}
\end{figure}

Typically, 2D fits to a galaxy velocity field assume that \vrad\ and
\vtan\ are constant along rings (e.g. the tilted ring analysis
described in \citet{Begeman89}).  By contrast, as mentioned earlier,
\vrad\ and \vtan\ do vary with azimuth, and indeed inspection of
Figures \ref{COarm} and \ref{Haarm} indicates that the primary
variations are due to the flow through the spiral arms rather than
variations with galactocentric radius.  As in most galaxies, the
rotation curve of M51 is relatively flat; the radial variations that
do occur are associated with spiral arm streaming.  Thus, we are
motivated to assume that radial variations of azimuthally-averaged
quantities are negligible (at least over relatively limited radial
ranges) and that \vrad\ and \vtan\ vary primarily with spiral arm
phase $\psi$.  The left panel of Figure \ref{armphase} shows the
relevant geometry depicting the spiral arm phase.  Our assumption is
that \vrad\ and \vtan\ are constant along narrow spiral arcs, such as
the segments in Figure \ref{armphase}, that are congruent to the
spiral arms.  Thus, we rewrite equation (\ref{vobs}) (for a limited
range of radii) as
\begin{equation}
{V_{obs}=V_{sys}+[v_R(\psi)\sin(\theta-\theta_{MA})+v_\theta(\psi)\cos(\theta-\theta_{MA})]\sin i}.
\label{newvobs}
\end{equation}

\begin{figure}
\includegraphics*[angle=-90,scale=0.70]{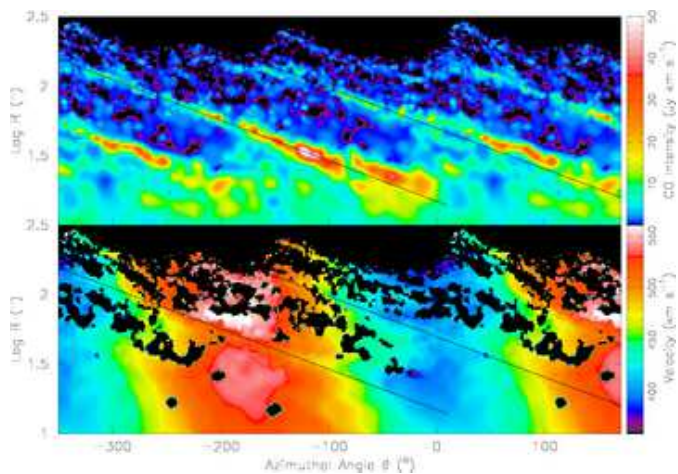}
\caption{\footnotesize Logarithmic polar projections of the CO
  intensity and velocity maps.  Though the origin of the abscissa
  (azimuthal angle) is arbitrary, in this case it is aligned with due
  North.  The direction of rotation is to the right (counter-clockwise
  as seen on the sky).  Also shown are the two logarithmic spiral
  lines positioned along the two spiral arms, which correspond to the
  lines overlaid on the maps of Figure \ref{COarm}.}
\label{pgCO}
\end{figure}

\begin{figure}
\includegraphics*[angle=-90,scale=0.70]{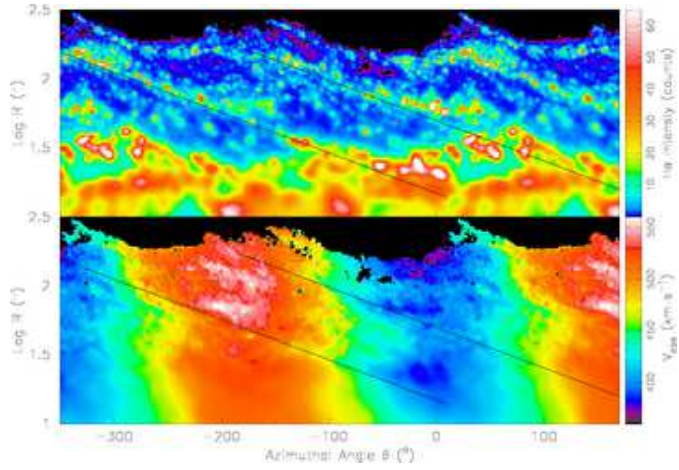}
\caption{\footnotesize Logarithmic polar projections of the H$\alpha$
  intensity and velocity maps.  Coordinate system and log spiral
  overlays are as in Figure \ref{pgCO}.}
\label{pgHa}
\end{figure}

In order to simplify the process of identifying regions of constant
arm phase, we adopt a coordinate system in which the spiral arms are
straight.  \citet{ElmSeidenElm89} show that the spiral arms of M51
appear as straight line segments in a $(\theta, log(R))$, or
logarithmic polar, coordinate system.  The right panel of Figure
\ref{armphase} shows the logarithmic polar diagram corresponding to
the features in the left panel.  Figure \ref{pgCO} shows the CO
intensity and velocity maps of M51 in log-polar coordinates, and
Figure \ref{pgHa} shows the corresponding H$\alpha$ maps.

The sky images in Figures \ref{COarm} and \ref{Haarm} are first
deprojected before being transformed into a $(\theta, log(R))$
coordinate system.  In order to deproject the sky view of a galaxy,
the center position, \pa, and \inc\ are required.  We initially use
the canonical values for these parameters, which are listed in Table
\ref{paramtab}.  We discuss the estimation of these parameters in the
next section.

The two straight lines overlaid on Figures \ref{pgCO} and \ref{pgHa}
indicate the adopted pitch angle of 21.1\degr\ and also correspond to
the spiral loci shown in Figure \ref{COarm}.  It is clear from the
overlaid lines, which are separated by 180\degr, that the weaker arm
is not symmetric with the brighter one, as discussed by
\citet{HenryQG03}.  Yet, both CO arms wrap around approximately
360\degr\ of the galaxy, even though they appear to jump in phase at
one or more positions.  The arms in H$\alpha$ show more jumps in phase
and variations in the pitch angle.  In spite of their asymmetries, the
CO spiral arms are particularly well described as logarithmic spirals,
better even than H$\alpha$, or the optical arms shown by
\citet{ElmSeidenElm89}.

We will refer to overlaid logarithmic spiral arcs (or lines) as
``slits,'' for we will extract observed CO and H$\alpha$ velocities as
a function of position along the arc (or line), similar to obtaining
long-slit spectra.  Each slit marks a region of constant arm phase
$\psi$.  Thus, while the observed velocity varies along the slit due
to projection, \vrad\ and \vtan\ are assumed constant.  We arbitrarily
define the arm phase marked by the leftmost slit in Figures \ref{pgCO}
and \ref{pgHa} as $\psi=0$\degr.  The other CO arm appears at an arm
phase of approximately $\psi$ = 180\degr; other features such as the
stellar arms and the gravitational potential minimum may of course be
offset from the CO arms.

As noted previously, our fit will assume that the intrinsic \vrad\ and
\vtan\ are constant at a given arm phase, i.e. along a given slit, but
that \vrad\ and \vtan\ vary with $\psi$ as the slit is translated in
azimuth.  Translating the slit amounts to shifting a straight line to
the right in the logarithmic polar diagram; this direction of
increasing azimuth is the same as the direction of rotation for M51.
We then fit equation (\ref{newvobs}) to the observed velocities
extracted at each arm phase $\psi$, thereby obtaining \vrad$(\psi)$
and \vtan$(\psi)$.

Although \vrad\ and \vtan\ vary primarily with arm phase, they may of
course also vary with radius.  Therefore, we limit the radial range of
an annulus (or equivalently the length of a slit) as much as possible
while still fitting a sufficiently extended azimuth range to obtain
good leverage on both \vrad\ and \vtan.  In other words, an annulus
should be sufficiently broad to cover the spiral arm both near the
major axis and the minor axis; the width of the annulus thus
depends on the pitch angle and galactocentric radius of the arm.

\begin{figure}
\epsscale{.8}
\plotone{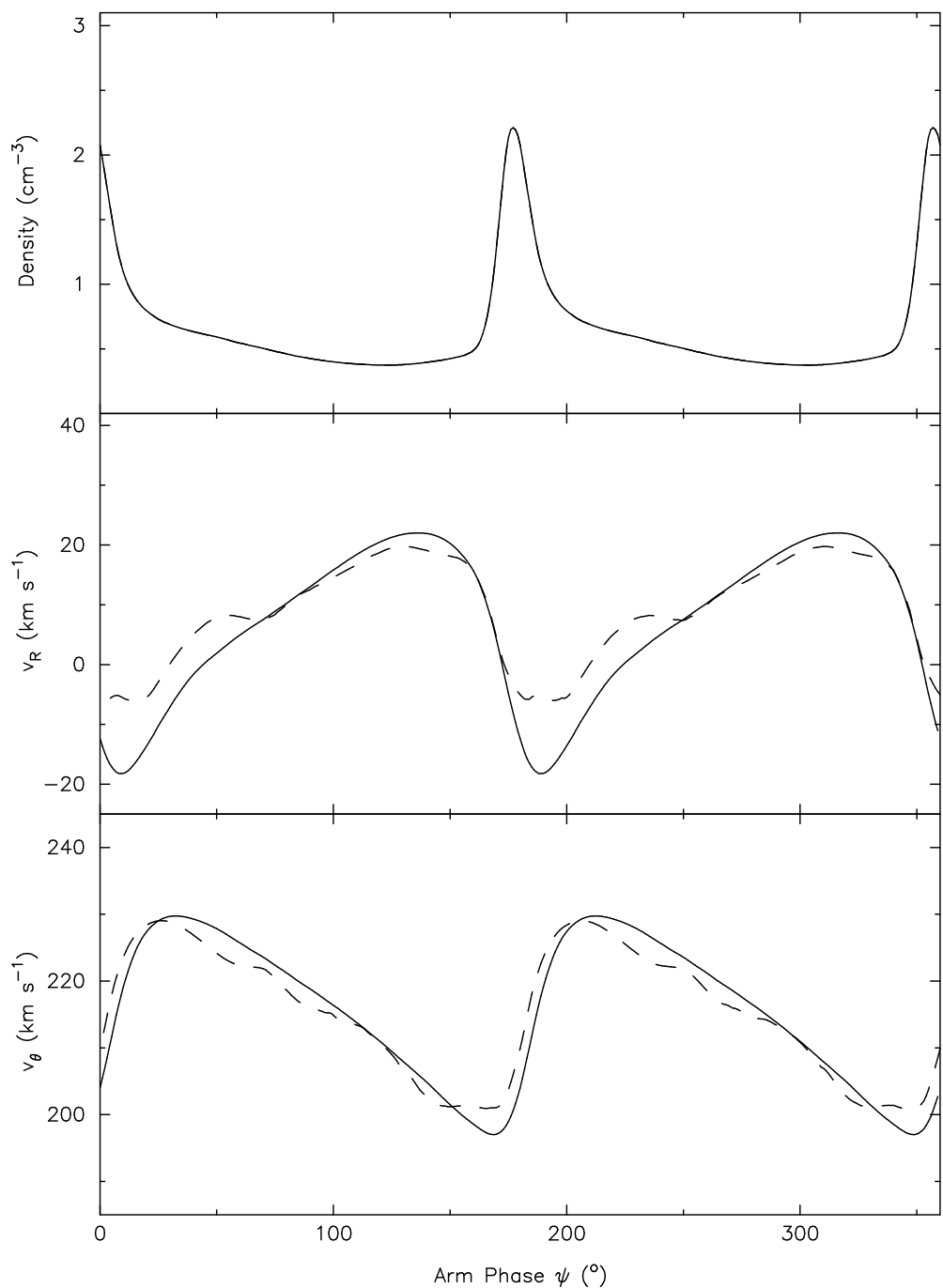}
\caption{\footnotesize Gas profiles as a function arm phase $\psi$ for
  a model spiral galaxy (see text).  Solid lines: density (top),
  \vrad\ (middle), and \vtan\ (bottom) profiles averaged at each arm
  phase, in an annulus extending from 8.38 - 8.92 kpc.  The \vrad\ and
  \vtan\ dashed lines in lower panels are obtained by fitting equation
  (\ref{newvobs}) to the ``observed'' velocities in the 8.38 - 8.92
  kpc annulus.}
\label{modvfit}
\end{figure}

We first test our method by applying it to a model spiral galaxy with
known radial and tangential velocities.  The solid lines in Figure
\ref{modvfit} shows the averaged density, \vrad, and \vtan\ profiles
in an annulus from a snapshot of a hydrodynamical simulation of a disk
responding to a spiral perturbation.  The model spiral galaxy is a 2D
version of a 3D model described in detail in \citet{GC02}, and the
annulus used here extends from 8.38 - 8.92 kpc.  The direction of gas
flow is in the direction of increasing phase.  As the gas approaches
the arm, (marked by density maxima) the radial velocity, \vrad,
decreases by $\sim$40 \kms.  The sign reversal of \vrad\ indicates
that the gas is moving away from the nucleus before the shock and
towards the nucleus after the shock.  As the gas emerges from the arm,
the radial velocity increases again.  The tangential velocity \vtan\
gradually decreases as the gas approaches the arm, then receives a
strong boost and reaches a maximum just downstream from the arm.

In order to test the fitting algorithm, the model \vrad\ and \vtan\ at
all locations are used in equation (\ref{vobs}) to create a model
observed velocity field.  This velocity field, along with the model
density map, are transformed into logarithmic polar projections.
Equation (\ref{newvobs}) is then fit to the model observed velocities
at each arm phase in an annulus, using slits parallel to the spiral
arms; the dashed lines in Figure \ref{modvfit} are the results of the
velocity fits, in the same annulus (8.38 - 8.92 kpc).  The results
reproduce the overall shape of the velocity profiles quite well,
although with slight phase shifts and offsets.  The offsets and the
shallower minimum in \vrad\ are likely due to the variation of the
pitch angle of the arms with radius; i.e.  the spiral arms are not
perfectly logarithmic, whereas the ``slit'' used to extract velocities
at constant arm phase is.  Despite these offsets, we were able to
reproduce the major features of the velocity profiles of the model
spiral galaxy, indicating that our method of fitting observed
velocities at each arm phase recovers a 2D velocity field reasonably
accurately.

\begin{figure}
\epsscale{1.2}
\plottwo{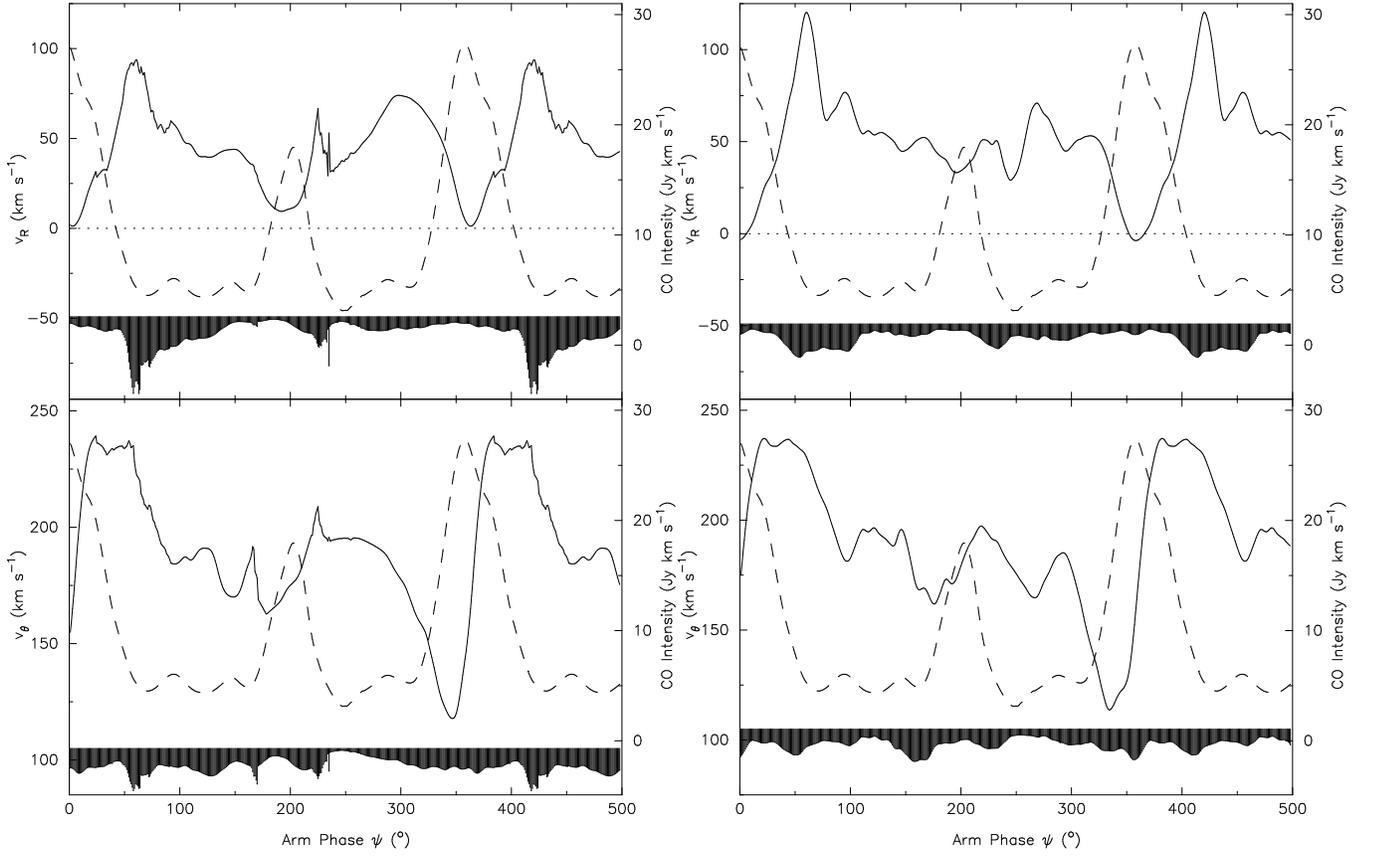}{f7b.eps}
\caption{\footnotesize CO (left) and \halpha\ (right) \vrad\ and
  \vtan\ fits as a function of arm phase for an annulus with an inner
  radius of 21\arcsec\ and an outer radius of 36\arcsec.  The one
  sided 3$\sigma$ error-bars are also shown on the bottom of each
  panel.  Dashed lines are the corresponding mean CO velocity
  integrated intensities, with the scale depicted on the right
  ordinate.  Table \ref{paramtab} shows the fixed (canonical)
  parameters used in obtaining these fits.}
\label{COvrvt}
\end{figure}

We now apply our fitting method to the M51 data, adopting systematic
parameters listed in Table \ref{paramtab}.  As an example, Figure
\ref{COvrvt} shows the \vrad\ (top) and \vtan\ (bottom) fits to the
observed CO and \halpha\ velocity field for one annulus between the
galactocentric radii of 21\arcsec\ and 36\arcsec. The CO intensity
averaged along a slit as a function of phase angle $\psi$ is also
shown as dashed lines, indicating the distribution of molecular gas.
As mentioned, $\psi = 0$\degr\ is arbitrary and is marked by the
leftmost line in Figures \ref{pgCO} and \ref{pgHa}, corresponding to
the brighter arm, which we will refer to as Arm 1.  We show a phase
range greater than 360\degr\ so that both upstream and downstream
velocities can easily be seen for both arms.  The direction of gas
flow through the arms (assuming we are inside corotation) is from left
to right, so that the right sides of the CO peaks correspond to the
downstream side of the arm.  In most cases, as the gas flows through
the arm, the radial velocity decreases and then increases, and the
tangential velocity receives a boost, as predicted qualitatively by
density wave theory.  

In a conventional tilted-ring velocity fitting analysis, galaxy
parameters such as the \inc, \pa, dynamical center, and \vsys\ can be
directly fit.  However, even though the \inc\ and \pa\ appear
explicitly in equation (\ref{newvobs}), for our fits all but the
\vsys\ must be assumed prior to deprojecting a galaxy velocity field
and therefore before the fit.  For our initial fits we employed the
standard assumed values for these parameters for M51, shown in Table
\ref{paramtab}.  In the next section we explore the effects of errors
in these assumed global parameters on the estimation of \vrad\ and
\vtan.

\section{Method Testing and Parameter Constraints \label
{methodtest}}

To test the sensitivity of the \vrad\ and \vtan\ fits to errors in the
global (fixed) parameters, we generated test velocity fields and
created sky projections with known parameters.  We then applied our
fitting technique to estimate \vrad\ and \vtan\ for the model
galaxies, assuming incorrect values for the fixed parameters, and
compared the fitting results to the actual model values of \vrad\ and
\vtan.  This enables us to quantify the sensitivity of the fits to the
fixed parameters.  In addition to constraints obtained from fitting
our kinematic data, we also use standard methods to constrain the
values of $V_{sys}$, $\theta_{MA}$, and $i$.  As we shall show, one of
our conclusions is that some of the basic parameters for M51, many of
which date to \citet{Tully74sp}, may in fact be poorly constrained due
to the morphological and kinematic perturbations induced by the tidal
interaction with its companion.

As an initial test, we generated a simple model with \vtan\ = 240
\kms\ and \vrad\ = $-$35, i.e. an axisymmetric disk with a flat
rotation curve and uniform radial inflow.  We refer to this model as
the ``constant velocity'' model.  We apply our general method to fit
\vrad\ and \vtan\ as a function of arm phase using ``observed''
velocities.  If we assume the input values of $V_{sys}$ = 464 \kms,
$\theta_{MA}$ = 170\degr, and $i$ = 20\degr, we indeed recover the
input values for \vrad\ and \vtan\ as independent of phase.  We now
consider the effects of assuming incorrect values for the parameters.

\subsection{Position of Dynamical Center \label{censec}}

In testing the sensitivity of the fits to the assumed center position,
we applied the fitting algorithm to a model for which the center
position was shifted by 1\arcsec\ in both RA and DEC.  We found that a
1\arcsec\ error in the assumed center has a negligible effect on the
fit velocities.  BIMA observations have an astrometric accuracy of
$\sim$10\% of the synthesized beam.  The highest resolution of our CO
observations is 4.5\arcsec, so the error in position will likely not
be greater than $\sim$0.5\arcsec.  Thus, observational errors will
likely not affect the results of our fits.  For all the analysis that
follows, we will adopt the center position listed in Table
\ref{paramtab}.  This choice assumes that the dynamical center
coincides with the location of a weak AGN known to exist in the
nucleus of M51 \citep{Hoetal87, NakaiKasuga88}.  We use the position
of the radio continuum source observed with the VLA, which has an
accuracy of $\pm$0$\farcs$01 \citep{Hagiwara01}.

\subsection{Systematic Velocity \label{vsyssec}}

Before discussing methods for determining $V_{sys}$, we first explore
the effect that an error in an assumed value of $V_{sys}$ would have
on fits for \vrad\ and \vtan\ in which $V_{sys}$ is held fixed, using
the constant velocity model.  As expected, an error $\Delta V_{sys}$
in the assumed $V_{sys}$ produces a sinusoidal variation in both
fitted velocity components, with an amplitude of $\Delta V_{sys}/\sin
i$, and a period of 360\degr (see eqn. \ref{vobs}).  Clearly,
$V_{sys}$ needs to be well determined.

\begin{figure}
\includegraphics*[angle=-90,scale=0.50]{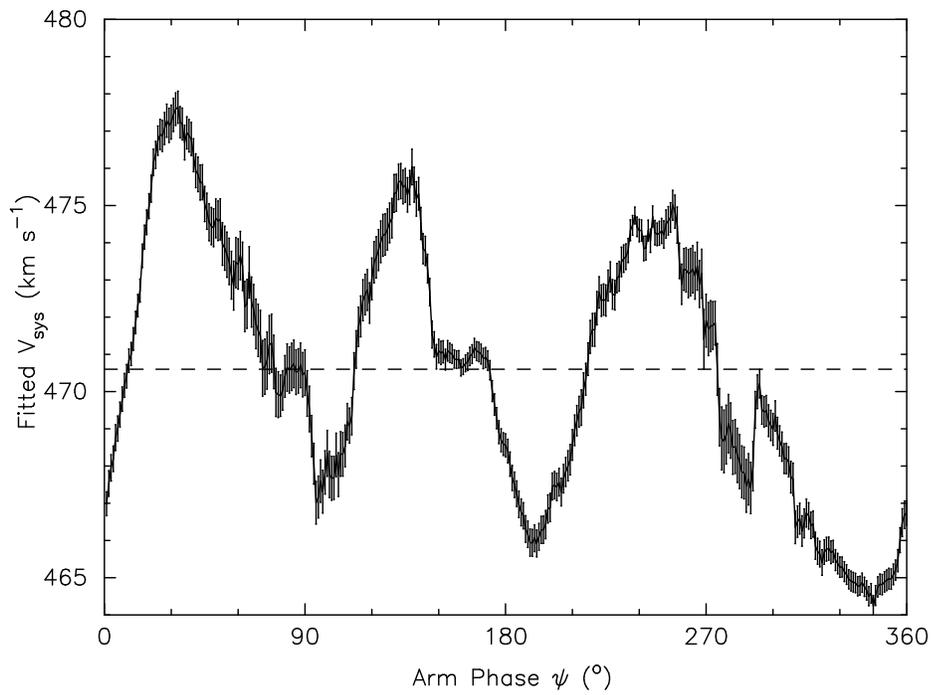}
\caption{\footnotesize Result of fit to M51 CO velocity data in which
  $V_{sys}$, \vrad, and \vtan\ were allowed to vary.  The radial range
  of the annulus is 14\arcsec\ - 136\arcsec.  The fact that the fit
  value of $V_{sys}$ varies with phase $\psi$ shows that other
  parameters (e.g. $i$, $\theta_{MA}$) vary with radius within the
  annulus.  The mean of the fits is 470.6 \kms\ (LSR), shown by the
  dashed line.}
\label{M51vsys}
\end{figure}

One approach to obtaining $V_{sys}$ is to fit the data for its value
using equation (\ref{newvobs}).  Figure \ref{M51vsys} shows the
results of fits to the M51 data in which $V_{sys}$ was fit, along with
\vrad\ and \vtan, as a function of arm phase.  Although $V_{sys}$
should be constant, it can be seen that the fit value of $V_{sys}$
varies with arm phase.  Similar variations in fitted $V_{sys}$ result
regardless of what values of the \pa\ and \inc\ are assumed.  

One possible explanation for the apparent variation of $V_{sys}$ is
that the galactic disk of M51 is twisted and/or warped, i.e.  the \pa\
and/or \inc\ may vary with radius.  We therefore compare the results
of fitting for $V_{sys}$ from a model galaxy with no warp to a model
with a warp.  We again make use of the constant velocity model, for
this model also represents an unwarped disk.  Instead of keeping
$V_{sys}$ fixed, we allowed this parameter to be free in the fit.  If
we use the true \pa\ and \inc, we correctly recover the adopted values
for all three free parameters, $V_{sys}$, \vrad, and \vtan.

\begin{figure}
\includegraphics*[angle=-90,scale=0.50]{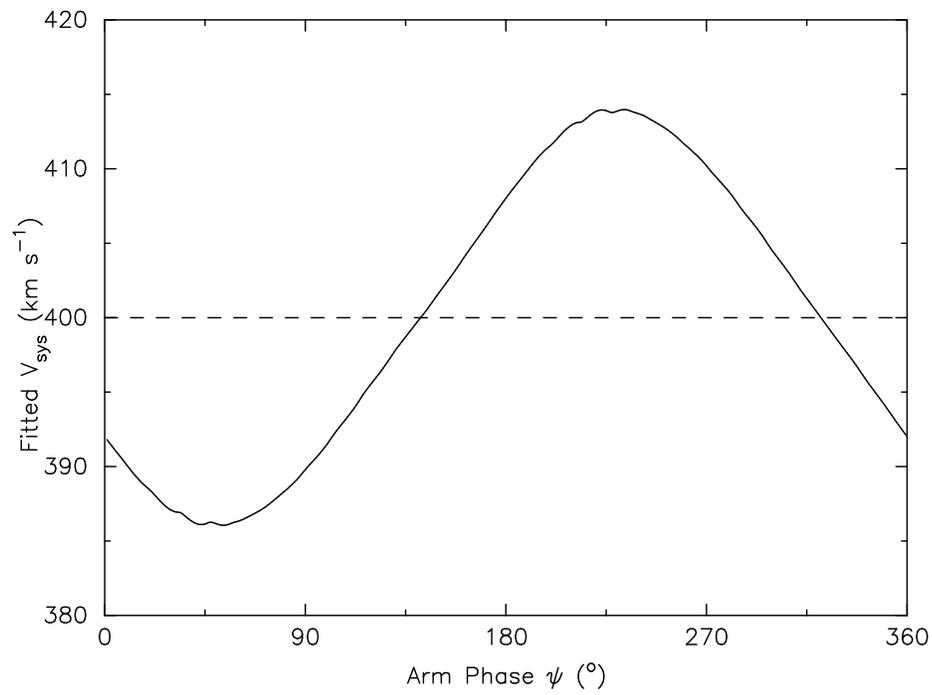}
\caption{\footnotesize Fit $V_{sys}$ as a function of arm phase,
  similar to Fig. \ref{M51vsys}, but for a model galaxy with a warp.
  The \vsys\ adopted for the model is 400 \kms\ (dashed line), equal
  to the mean of the fits (solid line).}
\label{warp}
\end{figure}

To generate a warp model, we increase $i$ monotonically from 25\degr\
at the inner radius (100\arcsec) to 35\degr\ at the outer radius
(200\arcsec); the \pa\ is kept fixed.  If we allow $V_{sys}$ to be
free in fitting the model data, Figure \ref{warp} shows that the
$V_{sys}$ varies almost sinusoidally about the true model value 400
\kms.  The mean fitted \vsys\ is equal to this value.  This is the
case regardless of what values of \inc\ or \pa\ we use (within
reasonable limits), and regardless of the radius range of the annulus
used for fitting.  Thus, regardless of the assumed fixed parameters
and of the limits in radius, for a simple warp the mean value of the
fits gives the correct \vsys.

Motivated by our finding that even with a warp the average fit value
of $V_{sys}$ gives the true value, we calculated the mean of the
$V_{sys}$ values shown in Figure \ref{M51vsys}, obtaining $V_{sys}$ =
470.6 \kms.  Comparison of the $V_{sys}$ fits of the actual M51 data
(Figure \ref{M51vsys}) with $V_{sys}$ fits to the simple warp model
(Figure \ref{warp}) shows that the warp model has a slower variation.
Hence, if a warp is responsible for producing variations in the fitted
\vsys\, it must be more complex than our simple model; we return to
this question in $\S$\ref{consmass}.  

We therefore apply two additional methods to estimate the value of
$V_{sys}$.  The first method is based on a standard tilted-ring
analysis \citep{Begeman89}, in which the galactic disk is represented
as a series of nested tilted rings.  In its most general form, each
tilted ring may have a different center, \vsys, \pa, \inc, and
rotational velocity.  We use 10\arcsec\ rings from an inner radius of
20\arcsec\ to an outer radius of 120\arcsec, fixing the center
position, \inc, and \pa\ to the values in Table \ref{paramtab}.  We
obtain mean \vsys\ of 471.4 $\pm$ 0.5 \kms.  When we allow the \pa\
to vary as well, we obtain a mean of 471.3 $\pm$ 0.3 \kms.

A third method we use to constrain $V_{sys}$ is to assume a functional
form for the rotation curve, using the {\tt NEMO} program {\tt
  rotcurshape} \citep{Teuben95}.  In contrast to the tilted-ring
method, which fits each ring independently, {\tt rotcurshape} fits
$V_{sys}$, $\theta_{MA}$, $i$, center position $\alpha$ and $\delta$,
and the coefficients of the function used to describe the rotation
curve simultaneously to the entire velocity field.  Therefore, it can
yield a single $V_{sys}$ that best fits the entire velocity field.  It
is particularly useful for finding $V_{sys}$ if the kinematic center
position can be fixed.  For this fit, we limit the {\tt rotcurshape}
fit to the inner 20\arcsec\ in radius.  This is inside the main spiral
arms, in the region where the rotation curve is rising and the
isovelocity contours are relatively straight.  We assume the center,
$\theta_{MA}$, and $i$ listed in Table \ref{paramtab} and a rotation
curve of the form $v = V_{sys} + v_o x/(1+x)$ where $x$ is the ratio
of the radius to the core radius, and fit for $V_{sys}$, $v_o$, and
the core radius.  We obtain $V_{sys}$ = 473 $\pm$ 0.5 \kms.  If we
allow the \pa\ to vary as well, we obtain 473.2 $\pm$ 0.3 \kms.  It is
encouraging that this is within 2 \kms of $V_{sys}$ determined from
the other two methods even though this fit uses a different method and
fits an entirely different region (i.e. the inner 20\arcsec, inside the
main CO arms, as opposed to outside 20\arcsec).

The canonical value of $V_{sys}$ for M51 is 472 $\pm$ 3 \kms\
\citep[and references therein]{Tully74sp}.  Table \ref{vsystab} lists
Tully's value for $V_{sys}$ as well as the results from applying the
three techniques described above.  Our different methods give a mean
\vsys\ of 471.7 $\pm$ 0.3 \kms.  Henceforth, we will fix $V_{sys}$ to
be 472 \kms\ (LSR, corresponding to a heliocentric velocity of 464
\kms) in fitting the velocity field to estimate $v_R(\psi)$ and
$v_\theta(\psi)$.

\subsection{Position Angle \label{pasec}}

To investigate the effect of errors in the assumed galaxy \pa\ on the
fitted values of \vrad\ and \vtan, we first use the aforementioned
constant velocity model.  Note that the \pa\ is required to deproject
the galaxy image, as well as in equation (\ref{newvobs}).  Using
incorrect \pa s, but correct model values of \vsys\ and \inc, yields a
greater effect on \vrad\ than on \vtan.  This is because an error
$\Delta\theta_{MA} \ll 1 $ in $\theta_{MA}$ results in an error
$\approx -v_{\theta}\sin\Delta\theta_{MA}$ in the fitted \vrad,
whereas the corresponding error in \vtan\ is $\approx
+v_R\sin\Delta\theta_{MA}$.  Since \vtan\ is large compared to \vrad,
the shift in \vrad\ is larger than the shift in \vtan.  Thus , a
$\pm$10\degr\ error in \pa\ in the ``constant velocity'' model which
has \vtan\ = 240 \kms\ produces approximately a $\mp$40 \kms\ shift in
the fitted radial velocity.  Position angle errors also produce small
perturbations in both velocity components.  We conclude that using an
accurate value of $\theta_{MA}$ is very important to obtain accurate
\vtan\ and especially \vrad\ fits.

Unfortunately, the \pa\ of the major axis of M51 is particularly
difficult to determine.  The strong spiral arms and the tidal
interaction with NGC 5195 distort the stellar disk, making it
effectively impossible to determine a \pa\ from the orientation of the
isophotes.  Thus, it is necessary to go beyond morphology in
determining $\theta_{MA}$.  We therefore revisit the determination of
the galaxy's position angle.  We apply the method of \citet{Tully74sp}
to our velocity data from H$\alpha$ and CO observations.  We also
study the effect of streaming motions on this method by using model
galaxies with known position angles and streaming velocities.  In
addition, we also apply two alternate methods to derive the position
angle.

The widely used value for the position angle of M51, 170\degr , was
determined by \citet{Tully74sp} using kinematic information.  Tully
assumed that the observed velocity should reach its extreme value at
the \pa\ of the galaxy major axis, $\theta_{MA}$.  To determine
$\theta_{MA}$, he averaged the observed velocities in wedges extending
over 5\degr\ in azimuth, and then for each radius took the \pa\ of the
wedge with the extreme velocity as the estimated major axis at that
radius.  Tully excluded radii at which he was not confident that the
true major axis had measured velocities (e.g. the faint interarm
regions near the major axis).

\begin{figure}
\epsscale{0.77}
\plotone{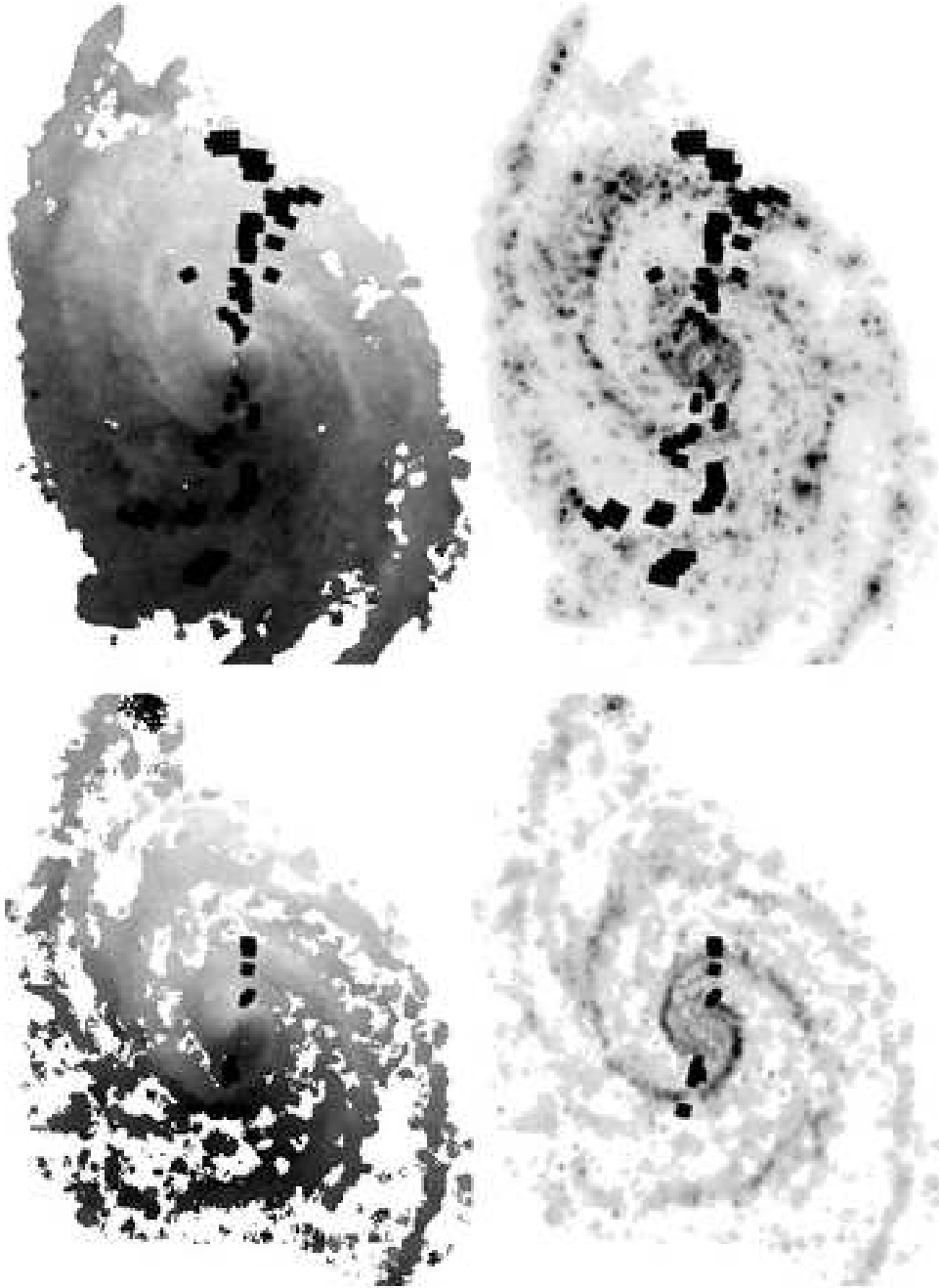}
\caption{\footnotesize Tully ``wedge'' method for estimating galaxy
  \pa.  The extreme velocity for each 5\arcsec\ annulus, averaged in
  5\degr\ wedges, is marked.  The upper panels show the \halpha\
  velocity (left) and velocity-integrated intensity (right); lower
  panels show the same for CO.  For CO, emission was too weak to apply
  the method at some radii, especially in the outer galaxy.}
\label{wedges} 
\end{figure}

Figure \ref{wedges} shows the results of applying Tully's position
angle determination method to H$\alpha$ and CO observations.  For each
annulus of radial extent 5\arcsec, the position of the wedge with the
extreme velocity is marked.  Due to the lack of data in the outer
regions of the CO observations, only the wedges in the inner
70\arcsec\ provide reliable measures of the \pa\ of the extreme
velocity.  Similar to Tully, we did not attempt to estimate the \pa\
of the extreme velocity at radii for which data are sparse in the
range of plausible \pa s of the major axis.  From the location of
these extreme velocity wedges, we found $\theta_{MA}$ to be 172\degr\
(from an error weighted average) from both CO and \halpha\
observations.

\begin{figure} 
\plotone{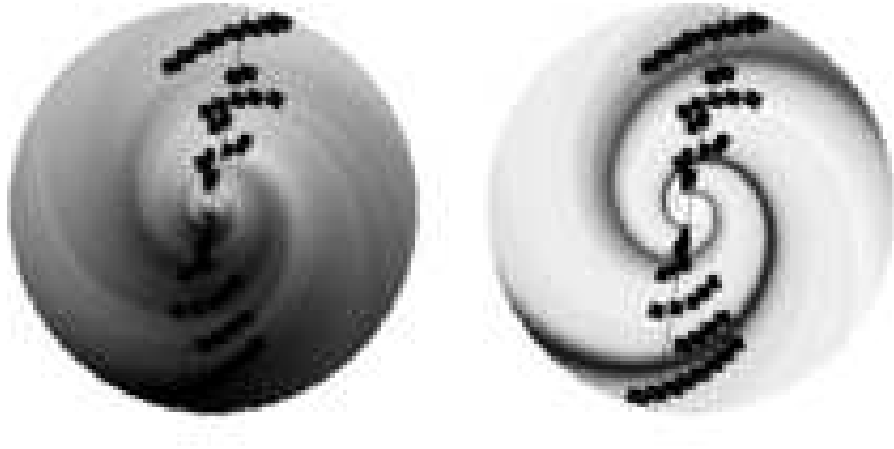}
\caption{\footnotesize Application of Tully wedge method to a model
  with streaming.  The velocity field is shown (left) along with the
  corresponding intensity map (right).  The position angle assumed in
  the model is 170\degr, shown by the solid line.  The 5\degr\ wedges
  with the extreme velocity for each annulus is marked.  It can be
  seen that streaming shifts the estimated \pa\ from the true \pa.
  Note that the extreme velocities do not occur in the arm.}
\label{modwedges}
\end{figure}

However, streaming motions can shift the velocities, resulting in the
extreme velocity occurring at position angles not corresponding to the
true major axis.  Indeed, inspection of positions of the velocity
extremum wedges overlaid on the intensity maps shows that the \pa s of
the wedges in the interarms are clearly shifted counter-clockwise from
those in the arms.  This is most evident in the H$\alpha$ maps, for
which emission is detected from almost everywhere in the disk.  We
further explore streaming effects on the Tully method using a model
with known streaming motions, generated using one of our \vrad\ and
\vtan\ fits to the M51 CO data, with $\theta_{MA}$ = 170\degr.  (Since
this test is designed simply to reveal the twists in the apparent \pa\
due to streaming, the particular value assumed for the true \pa\ and
the particular \vrad\ and \vtan\ fits used is not significant.)
Figure \ref{modwedges} shows the results of applying Tully's method to
this streaming model galaxy.  In any given annulus, the extreme
velocity averaged in the 5\degr\ wedges occurs in the interarm
regions.  In Tully's analysis, however, only spiral arm regions (near
the apparent major axis) were considered, due to observational
limitations.  Therefore, the major axis position angle he found is
likely biased clockwise from the true major axis.  As shown in Figure
\ref{modwedges}, even if the interarm regions are considered, the \pa
s of the locations of the extreme velocities do not necessarily
correspond to the major axis.  Thus, such an extrema method can be
biased due to the inherent streaming in M51, regardless of whether the
arms or interarms are considered.

\begin{figure}
\includegraphics*[angle=-90,scale=0.60]{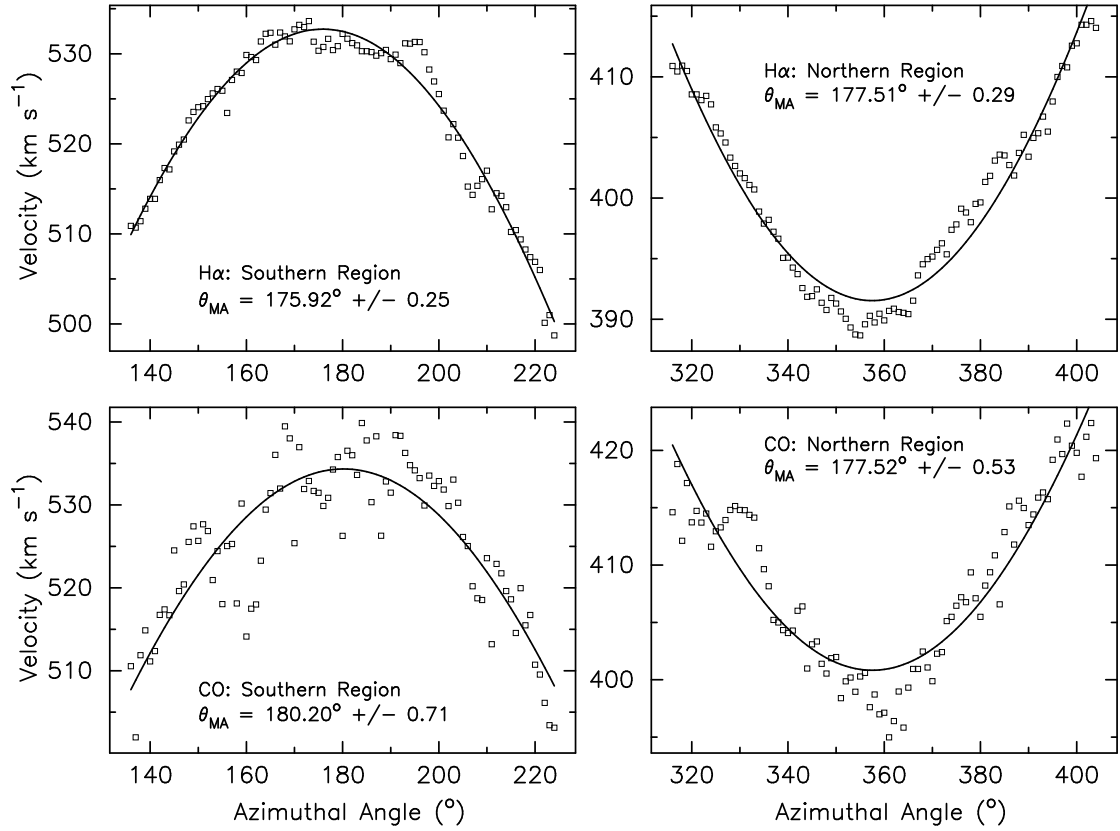}
\caption{\footnotesize Mean observed velocity plotted vs azimuthal
  angle.  All observed velocities were averaged over radius at each
  azimuth.  Velocities are fit by cosine functions (solid line);
  extremum of the cosine curve indicates the best fit \pa\ of the
  major axis.}
\label{radslits}
\end{figure}

We employ two alternate position angle determination methods that make
use of the full observed velocity field.  In the first method, we
average the observed velocity at each position angle in a wedge for
both the northern and southern sides of the galaxy; then we fit a
cosine curve to these averaged velocities as a function of azimuthal
angle.  We will refer to this method as the ``radial-averaged''
method.  This is most easily accomplished in the polar projection,
where we can average along a column to perform the radial average.
The radial-averaged velocity as a function of \pa\ is shown in Figure
\ref{radslits}, along with the corresponding cosine fitted curves.  We
assume that the galaxy major axis should be at the \pa\ of the extrema
of such curves.  The mean position angle determined from the H$\alpha$
and CO fits is $\sim$177\degr, larger than the position angle
determined by the Tully method.  Again, the position angle determined
in such a way is sensitive to streaming.  Earlier we showed that
streaming tends to cause the position angle of extreme interarm
velocities to be biased counter-clockwise from the true value.  Since
the interarms occupy a greater fraction of the galaxy compared to the
arms, streaming will introduce a counter-clockwise bias to the apparent
position angle of the major axis.  The effect of streaming on this
method is further discussed below, following a discussion of our
second position angle determination method.

\begin{figure}
\includegraphics*[angle=-90,scale=0.55]{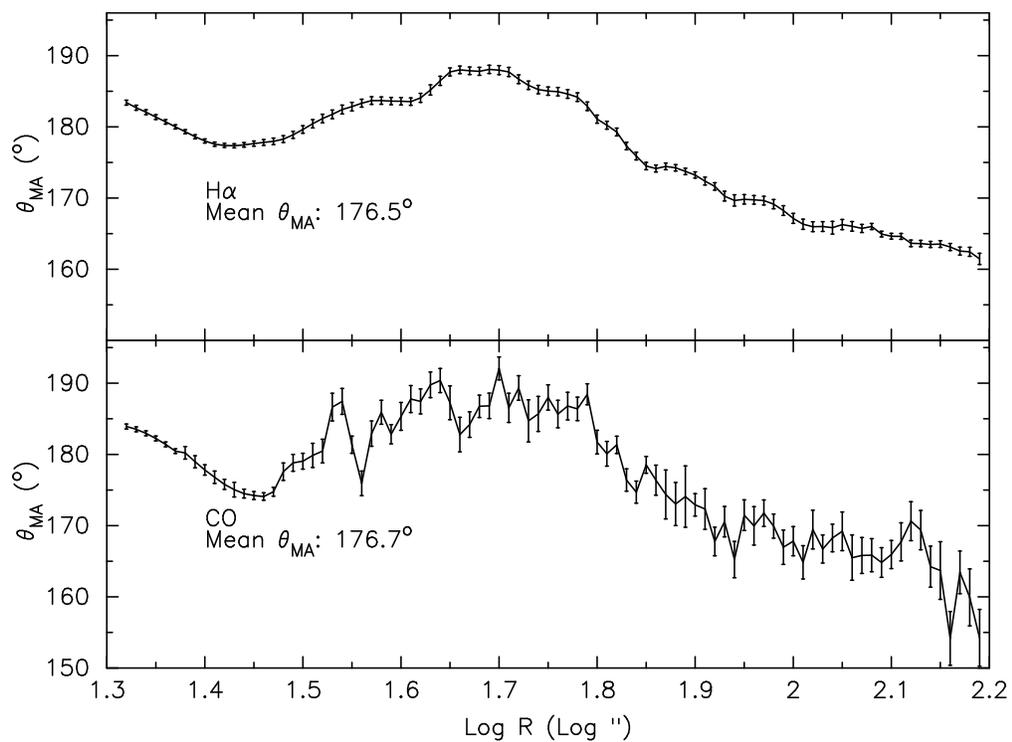}
\caption{\footnotesize Fit \pa\ of the major axis $\theta_{MA}$ as a
  function of galactocentric radius from \halpha\ (upper) and CO
  (lower) velocity fields.  The \pa s were obtained by fitting a
  cosine function to the distribution of observed velocity vs
  azimuthal angle at each radius.}
\label{circslits}
\end{figure}

Our second method to determine the \pa, the ``azimuthal fit'' method,
is similar to the one described in the previous paragraph, but instead
the cosine curve is fit to the observed velocities along a projected
circle with constant (projected) radius.  As in the previous method,
the polar projection of the velocity field is useful; in this case we
simply fit a cosine curve to the velocities along a row of constant
projected radius.  The results from applying this method are shown in
Figure \ref{circslits}.  Note that the \pa\ of the velocity extrema
varies systematically as a function of radius; it is approximately
180\degr\ in the inner region 30\arcsec\ from the center, declining to
165\degr\ 120\arcsec\ from the center; this trend, including the rise
near Log($R$) = 1.7, is also evident from simple inspection of Figures
\ref{pgCO} and \ref{pgHa}.  Averaging over the radius range displayed
in Figure \ref{circslits}, we obtain the same \pa\ of $\sim$177\degr\
as in the previous method.  Again, the velocities in the interarms
bias this determination of position angle, for the same reasons stated
in the previous paragraph.

In order to understand the effect of streaming motions on the \pa\ of
the major axis derived using the radial averaged and azimuthally fit
methods, we apply these methods to streaming models with known \pa s
and streaming velocities.  The model velocity fields are produced from
our \vrad\ and \vtan\ profiles obtained by assuming fixed values of
$\theta_{MA}$; we then apply the radial averaged and azimuthally fit
methods to these model velocity fields.  Both methods recover a
$\theta_{MA}$ of $\sim$176\degr, for all models, similar to the actual
M51 velocity field, even though the \pa s assumed in generating the
streaming models can be very different.  This is because for different
\pa s, the streaming velocities were derived from fits designed to
best match the observed velocities.  So in fact all the streaming
models give virtually identical observed velocity fields regardless of
assumed \pa.

However, if we recreate the models setting the radial component to be
zero everywhere, then we correctly recover the assumed position
angles.  This is clear evidence that radial streaming affects methods
to determine $\theta_{MA}$, not only near the minor axis, as recognized
by Tully, but also elsewhere including even the major axis.

In order to quantify the effect of the non-zero radial velocities on
the apparent position angle, we apply the position angle determination
methods to models with known constant radial velocities.  In these
artificial models tangential streaming velocities are assumed to vary
with arm phase, but radial streaming is assumed constant.  We found
that for every $\pm$10 \kms\ in radial velocity, the derived position
angle differs from the actual position angle by $\pm$ 3\degr.  This
degeneracy between the position angle and radial velocity renders it
difficult to accurately identify the true position angle, or to map
the radial velocity.  In order to accurately determine the position
angle, we need to know the radial streaming.  But in our effort to map
the radial and tangential velocities of M51, we need to know the
position angle.  Thus, as we carry out our investigation, we shall use
a range of position angles in deriving the two dimensional velocity
components of M51.

\subsection{Inclination \label{incsec}}

Estimating the inclination based on the orientation of the isophotes
is unreliable, as discussed in $\S$\ref{pasec}, due to the strong
perturbations from the spiral arms and the tidal interaction.  In
principle, the inclination can be determined from a fit to the
velocity field, as we did to obtain the systematic velocity in
$\S$\ref{vsyssec}.  However, the fit inclination is not well
determined by the available data, presumably due to the streaming.  To
test the sensitivity of the \vrad\ and \vtan\ fits to the inclination,
$i$, in equation (\ref{newvobs}), we assume incorrect values of the
\inc\ which differ from the true value by $\Delta i$ for the constant
velocity model (where \vrad\ and \vtan\ are constant).  We then fit
for \vrad\ and \vtan.  As expected, an error $\Delta i$ results in
errors in the fit value of \vrad\ and \vtan\ with magnitudes $\propto
\sin\Delta i$, along with small perturbations about this offset.

Since we find that error introduced in the velocity components due to
an incorrect \inc\ can be large (although just a simple scaling), we
sought other constraints on the inclination.  In particular, the
Tully-Fisher \citep{TullyFisher77} relation can be used to estimate
the inclination.  The well known Tully-Fisher relation is a
correlation between galaxy luminosity and maximum rotation speed.  The
inclination can be estimated by comparing the rotational velocity
predicted by the Tully-Fisher relation with the observed velocity of
the flat part of the rotation curve.  We use the baryonic form of the
Tully-Fisher relation discussed by \citet{McGaugh05} \citep[see
also][]{McGaugh00}:
\begin{equation}
{M_b=50\,V_c^4},
\label{Stacy}
\end{equation}
where $M_b$ is the baryonic mass (in $M_\odot$), and $V_c$ is the
circular rotational velocity (in \kms).  Since the dispersion in the
Tully-Fisher relation, $L \propto V_c^4$, is relatively small, given
the luminosity the uncertainty in $V_c$ is small.

In order to determine the baryonic mass $M_b$, we require the stellar
mass $M_*$, which is related to the $B$-band luminosity $L_B$ and the
$B$-band mass to light ratio $(M/L_B)$,
\begin{equation}
{M_{*}=L_B\cdot(M/L_B)}.
\label{Lb}
\end{equation}

We use the correlation of the galaxy color with $M/L$ discussed by
\citet{BelldeJong01} applicable to the \citet{CharlotBruzual91}
population synthesis models, to obtain $(M/L_B)$:
\begin{equation}
{(M/L_B)=10^{[-0.63+1.54(B-V)]}}.
\label{BC}
\end{equation}
The RC3 catalog \citep{RC3} gives $(B-V)=0.53$ for M51, so
$(M/L_B)$=1.54.

The last quantity required to determine $M_*$ is the luminosity $L_B$,
which can be derived if we know the distance.  Two independent studies
have given similar M51 distance estimates: observation of planetary
nebulae gives a distance modulus of $m-M$=29.62 $\pm$ 0.15
\citep{Feldmeier97}, and a study of surface brightness fluctuations in
the companion NGC 5195 gives $m-M$=29.59 $\pm$ 0.15 \citep{Jensen96}.
We thus employ a distance modulus of $m-M$= 29.6, corresponding to a
distance of $8.4 \pm\ 0.6$ Mpc.  Using the RC3 catalog value of
$B=8.67$, corrected for extinction,
\begin{equation}
{L_B=10^{-0.4(8.67-29.6-5.48)}L_\odot = 3.66\times10^{10} L_\odot}.
\label{lum}
\end{equation}
Using this value and the $(M/L_B)$ value of 1.54 (equation [\ref{BC}])
in equation (\ref{Lb}), we obtain
\begin{equation}
{M_*=5.64\times10^{10} M_\odot}.
\end{equation}

We can now apply the \citet{McGaugh05} relation in equation
(\ref{Stacy}) to obtain the circular velocity:
\begin{equation}
{V_c=[(M_* + M_{gas})/50]^{1/4}=188\,\,{\rm km\, s}^{-1}},
\label{Vc}
\end{equation}
where $M_{gas}$ is the total gas mass; in the case of M51 the gas is
predominantly molecular.  From our CO observations, we compute
$M_{gas} = 5.4\times10^{9} M_\odot$, using an X-factor of of $2\times
10^{20}$ cm$^{-2}$ [K \kms]$^{-1}$ \citep[e.g.][]{Strong88}.  Due to
the small value of the exponent in equation \ref{Vc}, errors in the
mass, due to variations in the X-factor, for example, will not
significantly affect the resulting rotational velocity.

The observed velocity is related to the circular velocity by
\begin{equation}
{V_{c,obs}=V_c\,\sin i}.
\label{sin}
\end{equation}
Adopting the center, $\theta_{MA}$, and $V_{sys}$ described in this
section, we apply a tilted-ring analysis to determine the flat part of
the rotation curve.  We obtain an observed circular velocity between
70 and 80 km s$^{-1}$, implying
\begin{equation}
{22^\circ\,\,\lesssim\,\,i\,\,\lesssim\,\,25^\circ}
\label{ninc}
\end{equation}
Therefore, for our subsequent fits, we adopt an inclination of
24$^\circ$.\footnote{Employing the standard Tully-Fisher relation
  instead, we obtain a mean inclination of $\sim$23\degr, using the slope
  and zero-point fits from \citet{Verheijen01}.}

\subsection{Summary: System Parameter Values}

In summary, we have shown that the fit values of \vrad\ and \vtan\ are
sensitive to the assumed values for the fixed parameters in equation
(\ref{newvobs}), $V_{sys}$, $\theta_{MA}$, and $i$.  Uncertainties in
the assumed position of the dynamical center are too small to
significantly affect the derived streaming velocities.  We have used
three different methods to determine $V_{sys}$, which resulted in a
value similar to the $V_{sys}$ found by Tully.  We have found it to be
extremely difficult to constrain the value of the \pa\ of the major
axis using the velocity field, due to the significant streaming that
shifts the position angle of the extreme velocities.  As a result, in
fitting for \vrad\ and \vtan, we allow for a range of plausible \pa s.
Lastly, we will adopt an \inc\ of 24\degr, which is determined by
using the baryonic Tully-Fisher relation \citep{McGaugh00} between the
baryonic mass and rotational velocity.  To estimate \vtan\ and \vrad,
we thus use the center position and systematic velocity listed in
Table \ref{paramtab}, but use a range of \pa s and an \inc\ of
24\degr.

\section{Results: Velocity Profile Fits \label{results}}

With our improved estimates of the global parameters, we apply the
fitting algorithm to the observed velocity field in different annuli
to determine the radial and tangential velocities \vrad\ and \vtan\ as
a function of arm phase $\psi$.  We initially adopt a \pa\
$\theta_{MA}$ of 170\degr.  We address the issue of a varying
$\theta_{MA}$ in $\S$\ref{consmass}.  Figures \ref{COvans} and
\ref{Havans} show the CO and H$\alpha$ \vrad\ and \vtan\ fits in 6
overlapping annuli between a galactocentric radii of 21\arcsec\ and
105\arcsec, and Figure \ref{anreg} shows the corresponding overlapping
annular regions.

An initial inspection of the streaming profiles indicates that the
velocity structure is rather complex.  Models of density wave
streaming qualitatively predict that as gas encounters the arm, \vrad,
which was positive (i.e. outward) in the interarm, becomes negative,
and as the gas exits the arm again becomes positive.  The azimuthal
velocity \vtan\ is predicted to increase rapidly as gas flows through
the arm, and then decline more gradually in the interarm
\citep[e.g.][and $\S$\ref{fitting}]{RobertsStew87}.  First we
concentrate on Arm 1 (the brighter arm, shown at $\psi$ = 0\degr\ and
more fully at 360\degr).  For \vrad\ there is a pronounced minimum
close to the arm position, seen in both CO and H$\alpha$.  There is a
boost in \vtan\ through the arm, again seen in both CO and H$\alpha$.
For this arm, the streaming is qualitatively as expected from
steady-state spiral shock models.  The velocities associated with Arm
2 (located at $\psi$ $\sim$ 200\degr) however do not agree with simple
predictions.  For \vrad, a clear minimum is only apparent in the outer
annuli, and the boost in \vtan\ is weak or nonexistent.  In the
interarms, the structure appears somewhat more complex than the simple
model expectation of a relatively constant or slowly rising \vrad\ and
a slowly declining \vtan.  We suggest that Arm 1 matches simple theory
because its structure is simple, i.e. well described as a log spiral
of constant phase.  By contrast, for Arm 2 the CO distribution is not
as well described by a single log spiral segment.  Instead, it has
several segments with different pitch angles and jumps in phase; thus
the velocities associated with this arm is complex.

One explanation for the differences in the two arms, as discussed by
\citet{RixRieke93}, is that the spiral pattern in M51 is actually a
superposition of a strong spiral mode with a $m$=2 Fourier component
with weaker $m$=1 and $m$=3 components.  \citet{HenryQG03}, using the
spatial distribution of CO emission obtained from the BIMA CO map,
found such a scenario to be feasible by explaining the bright arm as
the result of constructive interference between the $m$=2 and $m$=3
components, and the weak arm the result of a destructive interference
between the two components.  There is evidence for interarm structure
possibly supporting such a multiple density wave component description
of the spiral arms, which would be expected to manifest itself in the
kinematics.  Recent Spitzer observations of M51 clearly show spiral
structure between the main CO arms; the Spitzer image and interarm
features are discussed in the next section.

There are also clear differences between different annuli.  For
example, in the 36\arcsec\ - 61\arcsec\ annuli, the \vrad\ increase
downstream from the weaker arm is much more pronounced than in the
21\arcsec\ - 36\arcsec\ annuli.  In addition, there is a \vrad\
decrease to as low as $< -$50 \kms\ in the arms of the outer regions,
which perhaps can be attributed to an incorrect choice of a fixed \pa\
for the disk (see $\S$\ref{pasec}).  There are also clear differences
in the \vtan\ gradients between different annuli.

By and large, similar velocity structure is apparent in both CO and
\halpha.  For example, in the 47\arcsec\ to 80\arcsec\ annulus, the
gradual rise in \vrad\ from -50 \kms\ at $\psi$=180\degr\ to 70 \kms\
at $\psi$=300\degr\ is shown in both tracers.  Further, there is a
strong \vrad\ peak at $\psi$=120\degr\ in the 27\arcsec\ to 47\arcsec\
annulus in both CO and \halpha; however, such pronounced local extrema
in the interarms are not expected in the theory for a single spiral
mode.  In general, the overall amplitude of the streaming and the
location of most features coincide, and regions in which the velocity
structure is somewhat different tend to be interarm regions where
little CO is detected.

Such similarities are not unexpected due to the dynamical coupling
between the different components.  CO, which traces the molecular
component, is dynamically cold, with a velocity dispersion of only 4 -
8 \kms.  Thus, the molecular component of the disk reacts strongly to
any perturbation, as evident in the strong \vrad\ and \vtan\ gradients
associated with the spiral arms.  The spiral arms compress the gas,
triggering star formation.  The newly formed hot O and B stars
subsequently ionize the surrounding gas, resulting in \halpha\
emission.  Due to the fact that much of the \halpha\ emission comes
from gas {\it near} the region of birth, observed \halpha\ velocities
will be similar to observed CO velocities.  There may also be a
diffuse ionized medium not closely associated with the O and B stars,
and this medium is likely not dynamically coupled with the molecular
gas.  However, as can be seen in Figures \ref{pgCO} and \ref{pgHa},
the brightest regions of \halpha\ emission occur just downstream from
the molecular spiral arms.  Further, the generally good agreement
between the velocity measurements from CO and \halpha\ observations
suggest that most of the ionized emission originates in gas associated
with star forming regions.  This similarity in velocity structure
derived from independent observations also gives confidence that the
fitted velocities are reliable and that the deviations from simple
theory, including interarm features, are real.

The profiles in Figures \ref{COvans} and \ref{Havans} qualitatively
agree with previous studies of streaming in M51 involving 1D cuts
along the major and minor axes \citep[e.g.][]{Rand93, Aalto99}.  The
gradients of the velocity profiles through the arms in different
annuli is in accordance with the conclusion of \citet{Aalto99}
supporting the presence of shocks in the arms from a qualitative
comparison of velocities along 1D cuts to streaming models of
\citet{RobertsStew87}.  In $\S$\ref{analys} we analyze the feasibility
of a steady or quasi-steady spiral pattern in M51, which has been a
working hypothesis for many analyses of the spiral arms of this
galaxy.

\subsection{Interarm Structure}

In estimating the radial and tangential velocity components, we fit
observed velocities along log-spiral segments.  The slope of the slit
is determined by the slope of the main CO arms on the logarithmic
polar projection, i.e.  the pitch angle of the arms.  Though the slope
of the CO arms, or at least the bright arm, is well defined, that
slope may not be appropriate for the interarms.  In other words,
velocity may not be constant along the interarm log-spiral segments
congruent to the main CO arms.

The recent Spitzer 8 $\mu$m image of M51 \citep{Calzetti05,
  Kennicutt03}, shows clear interarm features not seen in the CO map
due to the lower resolution of the CO observations.  Many of these
features are spurs (or feathers) which have been found to be
ubiquitous in grand design spirals \citep{LaVigneVogelOstriker06}.
These interarm features will also cause kinematic perturbations.  In
fact, close inspection reveals that interarm perturbations in the
velocity field of M51 coincide with strong interarm features apparent
in the 8 $\mu$m image.  Since the features have different pitch angles
from the main CO arms, we are likely smearing out these finer interarm
velocity perturbations.  As a result, the interarm velocity profiles we
have derived do not reveal the details of the velocity perturbations
associated with interarm substructure; in a detailed study of the 2D
velocity field between the main arms, the interarm structure would
need to be considered.

In the next section, we use the \vrad\ and \vtan\ fits to assess the
feasibility of the hypothesis of a quasi-steady pattern.  Again, our
fitting method is designed to reveal streaming solely associated with
the spiral arms, and does not capture smaller scale perturbations,
such as those associated with interarm features.  Both observations
\citep{Elm80,LaVigneVogelOstriker06} and numerical simulations
\citep{KO02,ShettyOstriker06} have shown that spurs and feathers are
associated with star formation, indicating that these features are not
long lasting.  The modal theory, hypothesizing quasi-stationary grand
design spiral structure, acknowledges that such smaller scale features
can be transient \citep[e.g.][]{BertinLin96}.  In this study, smearing
out the interarm perturbations likely does not affect the overall
conclusions we draw from the fitted velocity profiles.

\section{Tests of Conservation Laws \label{analys}}

\subsection{Conservation of Vortensity \label{vort_sec}}

For a flattened system, the conservation of mass and angular momentum
can be combined to yield
\begin{equation} {\frac{\partial }{\partial t}\left(\frac{\nabla
        \times {\bf v}_{inertial}}{\Sigma}\right) +
    {\bf v}_{inertial}\cdot \nabla\left(\frac{\nabla \times
        {\bf v}_{inertial}}{\Sigma}\right)=0},
\label{fllvort}
\end{equation}
where $\Sigma$ is the surface density, and ${\bf v}_{inertial}$ is the
velocity in the inertial frame.  Equation (\ref{fllvort}) states that
the vorticity per unit surface density, known as vortensity, is
conserved along streamlines.  For steady systems, the conservation of
vortensity can be simplified:
\begin{equation} {\frac{1}{\Sigma}\cdot \left( \frac{v_\theta}{R} +
      \frac{\partial v_\theta}{\partial R} - \frac{1}{R}\frac{\partial
        v_R}{\partial \theta} \right) } = constant,
\label{vorteqn}
\end{equation}
because the temporal term in equation (\ref{fllvort}) vanishes.  Even
if the flow is not steady, portions of the galaxy that originated in a
region of constant vortensity will still satisfy equation
\ref{vorteqn}.

In order to test whether equation (\ref{vorteqn}) is satisfied for the
gas in M51, we need the surface density $\Sigma$, which we can
estimate using the observed CO brightness to derive the corresponding
H$_2$ column density.\footnote{As we will describe in
  $\S$\ref{contsec}, HI can be neglected since the gas in M51 is
  mostly molecular in the region studied.}  Most studies suggest that
the relationship between CO and H$_2$ is reasonably linear, though the
conversion factor, known as the X-factor, is controversial.  In our
analysis, we will assume that CO is indeed a linear molecular tracer,
and employ an X-factor of $2\times 10^{20}$ cm$^{-2}$ [K \kms]$^{-1}$
\citep[e.g.][]{Strong88}.

We first test the vortensity condition from the velocity profiles
derived in the 47\arcsec-80\arcsec\ annulus, (see Figure \ref{COvans}
- \ref{anreg}).  We choose this particular annulus because \vtan\
variations are relatively smooth in both arm and interarm regions, and
are likely due primarily to spiral streaming.  This annulus clearly
shows the characteristic \vtan\ boost in the arm, and the more gradual
interarm decrease in \vtan.  We consider the fits derived from this
annulus assuming a $\theta_{MA}$ = 170\degr (see Fig.
\ref{circslits}).  As shown in $\S$\ref{pasec}, changes in
$\theta_{MA}$ affect \vtan\ only modestly.

For the first term in equation (\ref{vorteqn}), $v_\theta / R$, we use
the mean value of the tangential velocities fit in the given region.
To measure the radial gradient of the tangential velocity, which
appears in the second term, we use
\begin{equation}
  \left.\frac {\partial{}}{\partial R} \right|_\theta = \frac{1}{R \tan i_p}\frac{d}{d\psi},
\label{rphase}
\end{equation}
where $i_p$ is the pitch angle of the spiral arms, and we assume
negligible variation parallel to the arm.  We adopt a pitch angle of
21.1$\degr$ (which is also the slope of the ``slit'').  We fit
straight lines to the velocity profiles in order to approximate the
last two terms in equation (\ref{vorteqn}).  For the surface
densities, we use the peak value for the arm, and for the interarm we
use the value of $\Sigma$ at phase separated by 90$\degr$ from the
arm.  Again, we are assuming that CO directly traces the molecular
abundance.  Table \ref{vorttab} shows the vortensity values for the
47\arcsec-80\arcsec\ annulus, including the values of each of the
terms in equation (\ref{vorteqn}).

Table \ref{vorttab} shows that, within the errors, the arms and
Interarm 2 have consistent vortensity values.  The value for Interarm
1, however, is lower than in the other regions.  The lower value for
Interarm 1 can be inferred directly from the profile itself (Fig.
\ref{Havans}).  The tangential velocities in both arms are clearly
rising, and \vtan\ in Interarm 2 (downstream from Arm 2 at $\psi = 200
\degr$) is predominantly decreasing, suggesting that the spiral arms
have the most significant influence on the velocities in these
regions.  In Interarm 1 (downstream from Arm 1 at $\psi = 0\degr$),
however, there is more structure to the velocities, suggesting that
there are other sources of perturbations in addition to the spiral
arms.  We find that for most of the velocity profiles in Figure
\ref{COvans} and \ref{Havans} the vortensity values are consistent
between the two arms, within the errors.  However, in the interarms,
the vortensity values differ.  We find varying vortensity values in
all interarm regions except for Interarm 2 indicated in Table
\ref{vorttab}.  Overall, the agreement between vortensity in arm
regions in each annulus indicates either that a steady state depiction
of the vortensity is valid and there is very little radial migration
of gas, or else that in a given annulus much of the gas originated in
a region of constant vortensity and has been conserved along
streamlines as gas in a whole annulus flows inward or outward.

\subsection{Conservation of Mass \label{consmass}}

\subsubsection{Flux Weighted Average \vrad \label{fluxvrave}}

In the QSSS scenario, the spiral pattern --- as defined by its
amplitude, phase, and rotation rate --- would not change significantly
over the course of a few revolutions in a frame rotating along with
the spiral pattern \citep{Lindblad63, BertinLin96}.  Such a framework suggests that
on average any accretion of material into the arms should be balanced
by the same amount of material exiting downstream.  This condition
corresponds to conservation of mass for a steady state system; if this
condition holds it should be apparent in the variation of observed
velocities with spiral arm phase.

\begin{figure}
\epsscale{0.93}
\plotone{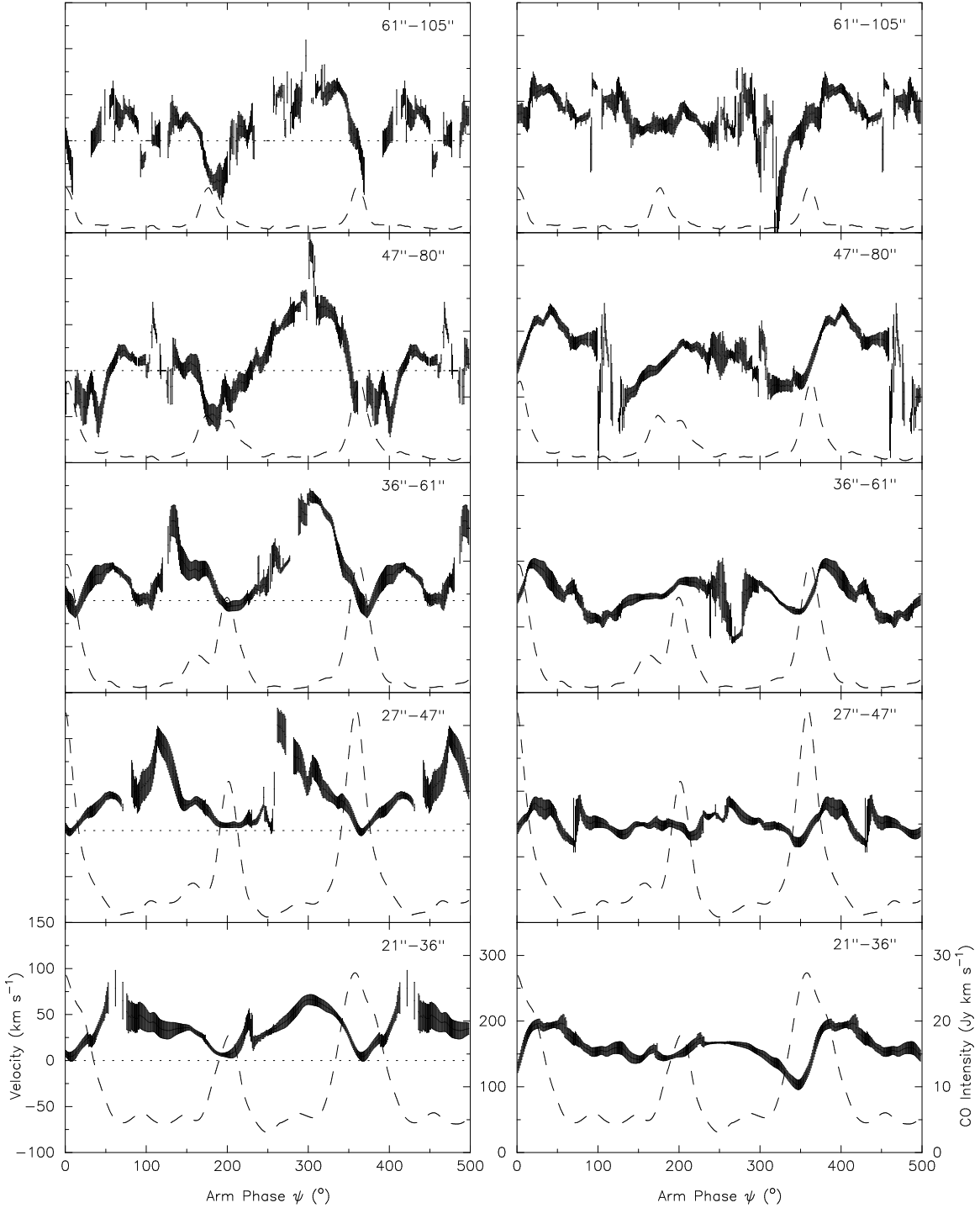}
\caption{\footnotesize \singlespace CO \vrad\ (left panels) and \vtan\
  (right panels) fits as a function of arm phase $\psi$ in different
  annuli (with radii labeled in the upper right of each panel).  The
  thickness of the line shows a range of $\pm3\sigma$.  Only \vrad\
  and \vtan\ fits with 3$\sigma$ $\leq$ 20 \kms\ and $\leq$ 60 \kms,
  respectively, are shown.  Dashed lines are the corresponding mean CO
  intensities, with the scale shown on the right ordinate.  We assume
  a \pa\ of 170\degr, an \inc\ of 24\degr\ and the center position and
  systematic velocity listed in Table \ref{paramtab}.  Figure
  \ref{anreg} shows the annular regions of M51 considered for these
  fits.}
\label{COvans}
\end{figure}

\begin{figure}
\epsscale{0.93}
\plotone{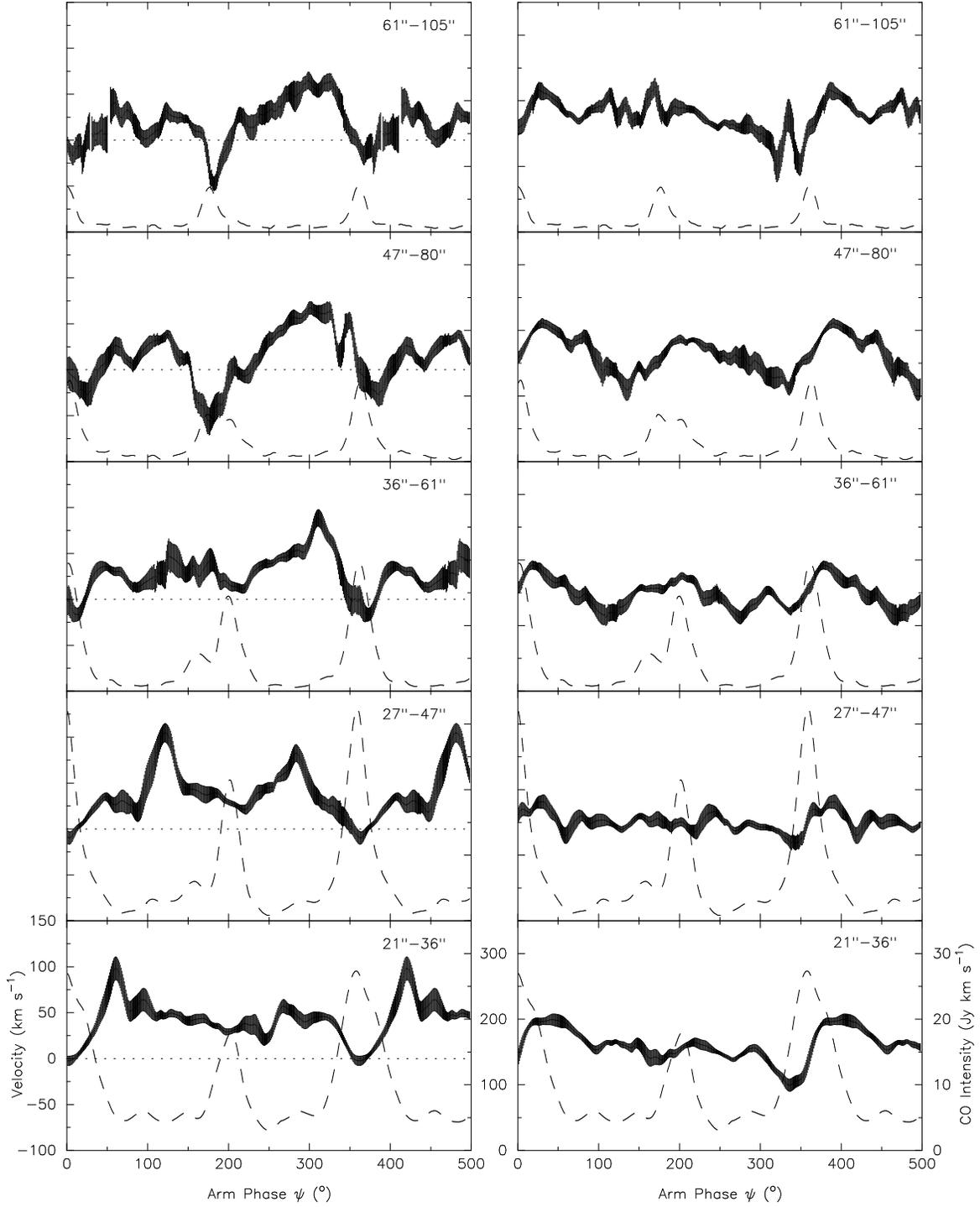}
\caption{\footnotesize H$\alpha$ \vrad\ and \vtan\ fits as a function
  of $\psi$ in different annuli, as in Figure \ref{COvans}.}
\label{Havans}
\end{figure}

\begin{figure}
\epsscale{.6}
\plotone{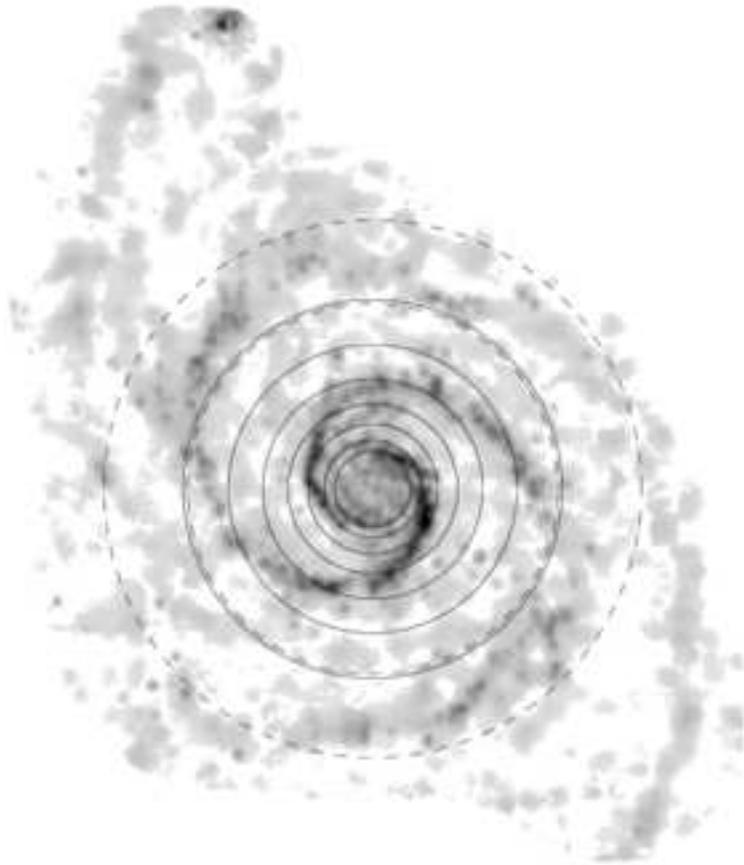}
\caption{\footnotesize Deprojected CO map of M51 showing the
  overlapping annuli for which \vrad\ and \vtan\ are fitted as a
  function of arm phase (shown in Figures \ref{COvans} and
  \ref{Havans}).  The radii of (solid) circles, from the inner to the
  outer, are: 21\arcsec, 27\arcsec, 36\arcsec, 47\arcsec, 61\arcsec,
  80\arcsec, and 105\arcsec.  The annulus marked by dashed circles
  (4.2 kpc $\le R \le$ 6.1 kpc) spans possible corotation
  radii corresponding to an adopted spiral pattern speed $\Omega_p$ =
  38 $\pm$ 7 \kms\ kpc$^{-1}$ (see $\S$\ref{contsec}).}
\label{anreg}
\end{figure}

As can be seen in Figures \ref{COvans} and \ref{Havans}, the spiral
arms clearly perturb \vtan\, with deviations of $\ge 100$ \kms\, but
\vtan\ always remains positive, indicating that the orbital flow is in
one direction only.  However, the radial velocities do change sign,
indicating both inflow and outflow.  If the spiral arms are indeed a
quasi-stationary pattern, then large amounts of matter should not be
undergoing net inflow or outflow, i.e. the mass-weighted average
radial velocity cannot be too large.  A large or spatially strongly
variable mass-weighted average \vrad\ would imply a very dynamic
system.  In particular, if the sign of this quantity changes, then
there would be a buildup or depletion of mass in one or more radial
locations.

As discussed in $\S$\ref{pasec}, the fit for \vrad\ is very sensitive
to the assumed value of the \pa.  Thus, in investigating the
mass-weighted average radial velocity, we consider a range of \pa s.
Figure \ref{pavr} shows the \vrad\ fits for \halpha\ in annuli with
radii of 47\arcsec\ - 80\arcsec\ and 61\arcsec\ - 105\arcsec\ for
three different \pa s, 170\degr, 175\degr, and 180\degr.  The CO fits
are similar, but noisier and have larger error bars (see Figs.
\ref{COvans} - \ref{Havans}).  A striking aspect of the fits in Figure
\ref{pavr} is the large magnitude of inflow in the arms for all three
\pa s; the radial velocity drops to as low as $-$75 \kms, suggesting
significant inflow for gas in the spiral arms.  In the upstream
regions \vrad\ is positive, approaching 70 \kms\ for some parameter
choices.  For a region farther in, in the 27\arcsec\ - 47\arcsec\
annulus shown in Figures \ref{COvans} and \ref{Havans}, the fitted
radial velocity (assuming $\theta_{MA}$ = 170$\degr$) reaches values
greater than 100 \kms, indicating tremendous outflow in the inner
regions; assuming a \pa\ of 180$\degr$ for this annulus only reduces
the peak velocity from $\approx$100 \kms\ to $\approx$75 \kms.

Figure \ref{flxvr} shows the flux-weighted average radial velocity,
$\langle v_R \rangle$, for different \pa s in the different annuli
used in the fitting process.  With the canonical \pa\ of 170\degr,
there is significant outflow in the inner regions of M51.  On the
other hand, a \pa\ of 180\degr\ seems appropriate for the innermost
regions of M51, since this yields a lower value of $\langle v_R
\rangle$.  However, with such a \pa\ we find significant inflow in the
outer region.  If we adopt an intermediate \pa\ of 175\degr, there is
outflow in the inner regions and inflow in the outer regions.  For a
$\theta_{MA}$ of 175\degr, $\langle v_R \rangle$ = 0 for the
36\arcsec\ - 61\arcsec\ annulus, with mean radius $\langle R \rangle$
= 1.98 kpc, while adjacent annuli have $\langle v_R \rangle$ = 10 and
$-$20 \kms, for $\langle R \rangle$ = 1.5 kpc and 2.6 kpc respectively
(1\arcsec\ = 40.7 pc at a distance of 8.4 Mpc).  If this were true,
then the gas would all collect near $R \approx$ 2 kpc in less than one
orbital time-scale ($\sim$200 Myr), which is not consistent with a
steady state.

This analysis leads us to conclude that if the spiral pattern is
long-lived, the large variations in the radial velocity shown in
Figure \ref{flxvr} suggests that the position angle must vary with
radius, indicating a disk which is not coplanar.  This trend
suggesting a larger \pa\ in the inner regions and a smaller \pa\ in the
outer regions is also in accordance with the \pa\ tests described in
$\S$\ref{pasec} (see Figure \ref{circslits}).

\begin{figure}
\includegraphics*[scale=0.70]{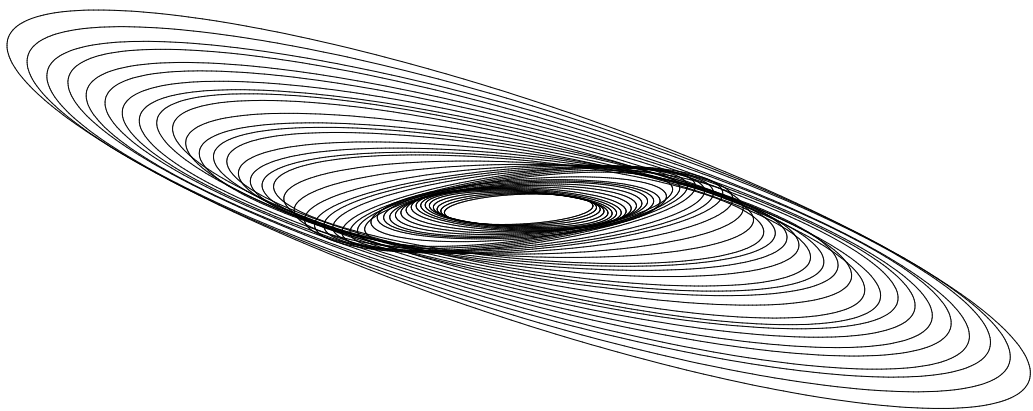}
\caption{\footnotesize Model disk showing the variation of the \pa\
  with radius.  The \pa\ profile is taken from Fig. \ref{circslits}.
  The inclination is exaggerated to show a more edge-on view.}
\label{epic}
\end{figure}

We schematically show a disk with a varying position in Figure
\ref{epic}; the \pa s of the ellipses are arranged as indicated by
Figure \ref{circslits}.  As discussed, one effect of a variation of
\pa\ is a disk that is not coplanar.  The inclination in this
schematic is exaggerated; the observed morphology, including the
apparent spiral structure, depends on the viewing angle, among other
factors.

\subsubsection{Continuity and Spiral Pattern Speed \label{contsec}}
In this section we explore the plausibility of QSSS using the gas
continuity equation.  The continuity equation for gas flow in a
two-dimensional system is 
\begin{equation}
\frac{\partial\Sigma}{\partial t} +
  \nabla\cdot(\Sigma{\bf v})=0;
\label{cont}
\end{equation}
this holds in any frame, e.g. whether the velocity is measured in an
inertial frame or one rotating at a constant pattern speed.  The first
term, $\partial \Sigma / \partial t$, represents the temporal growth
or decay of the surface density $\Sigma$ at any given radius $R$ and
azimuthal angle $\theta$ in the plane of the galaxy, where those
coordinates are with respect to the frame in which the velocity is
being measured.

\begin{figure}
\includegraphics*[angle=-90, scale=0.60]{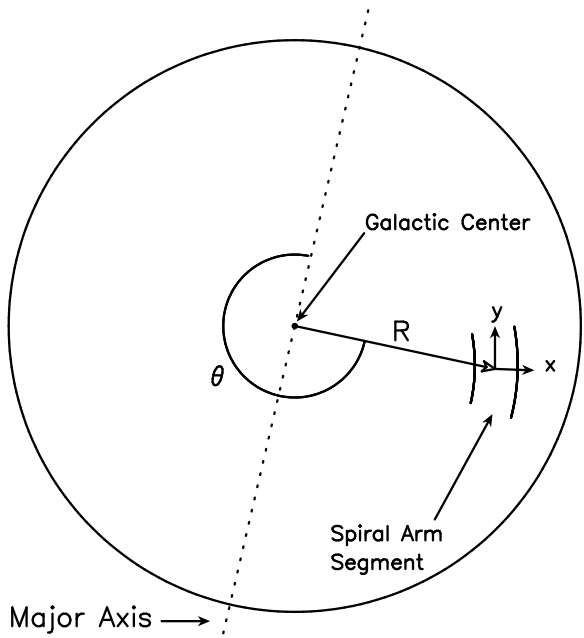}
\caption{\footnotesize Coordinate transformation geometry, from
  $(R,\theta)$ galactocentric coordinates to the $(x,y)$ spiral arm
  frame.}
\label{armgeom}
\end{figure}

If the flow is in a steady state, then the temporal term vanishes,
leaving only the mass flux term $\Sigma{\bf v}$.  If the gas is
responding primarily to a single dominant spiral perturbation, as would
be required for a fixed spiral pattern, and when ${\bf v}$ is measured in the frame
rotating at the pattern angular velocity $\Omega_p$, 
\begin{equation}
\nabla\cdot[\Sigma({\bf v}_{inertial}-\Omega_p R \hat{\theta})]=0.
\label{cont_anal}
\end{equation}
Thus, for an exact steady state the mass flux must be constant (in the
frame rotating with the same angular velocity as the spiral mode).
For a quasi-steady state, the temporal variations in $\Sigma$ will
only be small, and thus variations in mass flux would also be small.
This condition can be further simplified using a reference frame
aligned locally with the spiral arms.  Figure \ref{armgeom} shows this
reference frame; the $x$ and $y$ coordinates are the directions
perpendicular and parallel to the local spiral arm, respectively.  The
transformation between cylindrical coordinates and this arm frame is
achieved using
\begin{equation}
\hat{x}=\cos i_p \hat{R} + \sin i_p \hat{\theta},
\label{xtran}
\end{equation}
\begin{equation}
\hat{y}=-\sin i_p \hat{R} + \cos i_p \hat{\theta},
\label{ytran}
\end{equation}
where $i_p$ is the pitch angle of the arms.  

The velocities in the arm frame are given by
\begin{equation}
v_x=v_R \cos i_p + (v_\theta -\Omega_p R) \sin i_p ,
\label{vxeqn}
\end{equation}
and
\begin{equation}
v_y=-v_R \sin i_p + (v_\theta -\Omega_p R) \cos i_p .
\label{vyeqn}
\end{equation}

From the maps shown in Figures \ref{COarm} and \ref{pgCO}, it is
apparent that the intensity and velocity vary significantly more
across the arms, in the $x$-direction, than along them, in the
$y$-direction.  Thus, the variation in the product of $\Sigma v_y$
along $\hat{y}$ is much smaller than the variation in the product of
$\Sigma v_x$ along $\hat{x}$, reducing equation (\ref{cont_anal}) to
\begin{equation}
{\Sigma v_x\approx constant}.
\label{sigvx}
\end{equation}
Namely, for a steady pattern, as the gas decelerates (in the $x$
direction, perpendicular to the arm), mass accumulates and the surface
density increases; as the gas velocity increases, the surface density
decreases.

One difficulty in testing whether equation (\ref{sigvx}) holds is that
neither $\theta_{MA}$ nor $\Omega_p$ is well constrained.  Errors in
$\theta_{MA}$ yield errors in the fitted value of \vtan\ of $\approx
v_R \sin \Delta \theta_{MA}$ and in the fitted value of \vrad\ of
$\approx -v_\theta \sin \Delta \theta_{MA}$.  If $\Delta\Omega_p$ is
the error in the pattern speed, then the fitted value of $v_x$ will be
approximately given by $v_x + (-v_\theta \sin \Delta \theta_{MA} \cos
i_p + v_R \sin \Delta \theta_{MA}\sin i_p - \Delta \Omega_p R \sin
i_p)$.  Since the \vrad\ term has two factors of the $\sin$ of small
angles, that term will be much smaller compared with the other two
terms.  The true value of $v_x$ will therefore differ from the fitted
value by $C_x \approx v_\theta \sin \Delta \theta_{MA} \cos i_p +
\Delta \Omega_p R \sin i_p$.

In order to assess whether steady state continuity as expressed by
equation (\ref{sigvx}) holds in the case of M51, we therefore consider
the quantity
\begin{equation}
  \Sigma \tilde{v}_x=\Sigma (v_R \cos i_p + (v_\theta - \Omega_p R) \sin
  i_p + C),
\label{modvxeqn}
\end{equation}
where \vrad\ and \vtan\ are fitted values and $C \equiv \langle C_x
\rangle$, i.e. the (unknown) azimuthally-averaged correction due to
the errors in $\theta_{MA}$ and $\Omega_p$.  We apply equation
(\ref{modvxeqn}) by solving for the value of $C$ using the values of
\vrad, \vtan, and $\Sigma$ in the two arm segments of an annulus:
\begin{equation}
  C =  \frac{[v_{R,arm1} \cos i_p + (v_{\theta,arm1} - \Omega_p R) \sin i_p]
    \Sigma_{arm1} - [v_{R,arm2} \cos i_p + (v_{\theta,arm2} - \Omega_p R)
    \sin i_p] \Sigma_{arm2}}{\Sigma_{arm2} - \Sigma_{arm1}}.
\label{modvxC}
\end{equation}
We then test whether the value of $C$ obtained using equation
(\ref{modvxC}) also satisfies equation (\ref{modvxeqn}) in the
interarm regions.  If equation (\ref{modvxeqn}) is satisfied for both
interarm and arm regions, it would suggest an approximate steady
state.

We again focus on the 47\arcsec-80\arcsec\ annulus, where the \vrad\
(and \vtan\ ) are relatively ``smooth.''  The \vrad\ profiles for this
annulus is shown in Figure \ref{pavr}, assuming three different values
of the position angle.  Table \ref{mctab} shows the relevant values
associated with equation (\ref{modvxeqn}) for the arms and interarm
regions.  We employ the pattern speed of 38 $\pm$ 7 \kms\ kpc$^{-1}$,
calculated by \citet{ZimmerRandMcG04} by applying the
Tremaine-Weinburg method CO observations of M51.  Corotation
corresponding to this pattern speed is marked on Figure \ref{anreg}.
After solving for $C$ using quantities from the arms, it is clear that
the flow in the interarm region is not consistent with a steady state
description.  A \pa\ of 170$\degr$ produces large negative mass flux
in the arms, and positive flux in one interarm region.  Even
variations in the X-factor cannot resolve this discrepancy.  Assuming
larger values of the \pa\ still produces mass fluxes with vastly
different magnitudes, and even different signs.  Increasing the error
in $C$ up to an order of magnitude still cannot result in consistent
mass fluxes between the arm and interarm.  This suggests that any
reasonable changes to the values of $\theta_{MA}$ or $\Omega_p$ will
still result in varying mass fluxes.  We have checked the mass flux in
other annuli using the same method as for the 47\arcsec-80\arcsec\
annulus, as well as in other localized regions not presented here, and
found similar discrepancies in the mass flux.

In our analysis of continuity so far, we have not taken into account
the contribution from the atomic component of the disk.  In fact, in
most galaxies the majority of the gas exists in the form of HI.  In
M51, \citet{TilanusAllen89} showed that the downstream offset of HI
relative to the dust lanes is likely due to dissociation of molecular
gas by recently formed massive stars.  However, the inner disk of M51
has an unusually large fraction of molecular gas, so even at peaks of
the HI photodissociation arms, the contribution of atomic gas to the
total gas surface density is negligible.  Using the HI maps of
\citet{Rotsetal90}, we find that the atomic column density N(HI) is
significantly less than the molecular column density N(H$_2$) in the
vast majority of locations in the inner disk (21\arcsec\ $\le R \le
105$\arcsec); N(HI) exceeds N(H$_2$) in only $\sim$ 7\% of the inner
disk.  The mean value of N(H$_2$)/N(HI) throughout the inner region is
$\sim$10.  Though we used a constant X-factor to obtain the molecular
surface density, moderate variations in the X-factor \citep[for M51,
see][]{NakaiKuno95} will not be sufficient to account for the
discrepancy.  Nevertheless, no change in the X-factor, or in the
contributions of the molecular or atomic matter to the total mass, can
account for the change in sign of the mass flux; the varying sign of
$\Sigma v_x$ can only be due to a sign change in $v_x$, not $\Sigma$.

Our conclusion, after analyzing the mass flux, is that the kinematics
are not consistent with a quasi-steady spiral pattern in a flat disk.
We find that no single pattern speed can satisfy quasi-steady state
continuity, suggesting that the QSSS hypothesis is not applicable to
M51.  It is essentially the tremendous variations of the radial
velocity within a given annulus --- amounting to $\sim 100$ \kms\ ---
that lead the QSSS hypothesis into difficulty.  One explanation for
the transient nature of the spiral arms in M51, perhaps due to the
interaction with its companion, is a spiral perturbation with a
constant pattern speed, but with time-varying amplitude.  Or, there
may be multiple modes at work in the disk of M51, which may be
construed as a mode with a radially varying pattern speed
\citep[e.g.][]{MerrRandMeidt06}.  Multiple patterns speeds in M51 have
been previously suggested by \citet{Vogel93} and
\citet{ElmSeidenElm89}.  However, the extreme variations in the radial
velocity cannot be explained by multiple patterns alone.  Possible
causes for the large observed \vrad\ gradients are large out-of-plane
motions or a variation in inclination; since the inclination of M51 is
small, a variation in $i$ due to a warped or twisted disk will produce
large variations in the observed velocity due to projection effects.

\subsection{Discussion}

We have shown that the density and velocity structure in M51 does not
support a quasi-steady state depiction for the spiral pattern, using
measurements of the mass flux.  Further evidence that the observed
structure is inconsistent with steady state can be obtained by
adopting the fitted 2D velocity field, and demonstrating that the
density structure is then non-steady.  We have carried out this
exercise using a modified version of the {\tt NEMO} task {\tt
  FLOWCODE} \citep{Teuben95}.  In this exercise, a disk is populated
with gas tracer particles using the intensity profiles averaged along
spiral segments, reproducing the spiral density pattern of M51.  Each
location in the disk has an associated \vrad\ and \vtan, given by the
fitted velocity profiles (e.g. Figures \ref{COvans} and \ref{Havans}),
and an assumed value of the pattern speed.  The motion of the
particles is then integrated using {\tt FLOWCODE}: after a suitably
small timestep, the particles take on new velocities depending on
their location in the disk.  In essence, this simulation is a purely
kinematic test to determine whether the steady state continuity
equation (eqn.  [\ref{cont_anal}]) is satisfied or not, using the
density and fitted velocity profiles of M51 (Figs.
\ref{COvans}-\ref{Havans}).  We find that the input spiral pattern
vanishes in less than one orbital time scale ($\sim$200 Myr),
regardless of what values of the \pa\ and pattern speed we assume.

The precise nature of the velocities is one of a number of issues that
need to be considered in further studying the the global spiral
pattern in M51.  For example, our result suggests the role of a warp
certainly needs to be taken into account.  There are strong
indications that the outer disk of M51 is warped; our finding suggests
that the disk is not coplanar even further inward.  The non-coplanar
attribute may be the result of the tidal interaction between M51 and
its companion.

The possible warp and/or twist in the disk of M51 would of course
affect the projected velocities, and would present itself as gradients
in the velocity components, as discussed in the previous section.  If
this were indeed the case, then the single or multiple in-plane modes
would have to be in phase with the vertical mode in order to sustain a
spiral pattern.  The inherent uncertainty in deriving three velocity
components from the single observed component leads to difficulty in
estimating and analyzing both the vertical and in-plane modes.

\section{Summary\label{summary}}
We have analyzed the velocity field of M51, using CO and \halpha\
observations, to investigate the nature of the spiral structure.  We
summarize the main results here:

1) The velocity field is quite complex.  Observed velocities show
significant azimuthal streaming associated with the spiral arms, as
well as strong gradients in the radial velocities.

2) The aberrations in the velocity field strongly suggest that the
disk is not coplanar, perhaps as far in as 20\arcsec\ ($\sim$800 pc)
from the center.

3) We obtain fitted radial and tangential velocity profiles by
assuming that velocities in any annulus vary only with arm phase.
Strong gradients in the radial and tangential velocities are found in
the profile fits.  In general, the shape of both the \vrad\ and \vtan\
profiles are in qualitative agreement with theory of nonlinear density
waves, and support the presence of shocks.

4) In detail, the velocity profiles from different radial regions of
M51 differ significantly.  In addition, velocity profiles associated
with the two arms also show differences in a given annulus.  For the
arm that is well described by a logarithmic spiral (bright arm), the
associated velocities are in good agreement with simple theoretical
spiral shock profiles.  For the other arm, which is not as well
described by a logarithmic spiral, the velocities are more complex.

5) The velocity profile fits from CO and \halpha\ emission are rather
similar, suggesting that most of the \halpha\ emission originates from
gas associated with star forming regions.

6) When we assume a single value for the position angle of the major
axis of M51 and inclination, we find that large amounts of material
flows toward an annulus of intermediate radius, due to the large
gradients and change of sign in the flux weighted average radial
velocity.  As a result, either the position angle of the major axis or
the inclination must vary with radius, suggesting that the disk of M51
is warped and twisted.

7) We analyze conservation of vortensity, using the radial and
tangential velocity profile fits.  We find that vortensity is fairly
consistent within a given annulus, indicating that the gas there all
originated in a region of uniform vortensity.

8) Using the equation of continuity, we find that the density and
fitted velocity profiles are inconsistent with quasi-steady state mass
conservation in any frame rotating at a constant angular speed, at
least for a planar system.  Variations in the pattern speed, \pa, and
X-factor alone cannot account for the differences in the mass flux,
suggesting that spiral arms are quite dynamic, and possibly that
out-of-plane motions are significant.

\acknowledgements We are grateful to S. McGaugh for his help in using
the baryonic Tully-Fisher relation to estimate the inclination of M51.
We thank W.T. Kim for useful discussions, and G. Gomez for providing
velocity profiles from his hydrodynamic models.  We also thank the
anonymous referee for raising many interesting questions and useful
suggestions.  This research was supported in part by grants AST
02-28974 and AST 05-07315 from the National Science Foundation.

\bibliography{ref}

\clearpage

\begin{deluxetable}{lcc}
\tablecolumns{3}
\tablewidth{0pt}
\tablecaption{Initially Adopted Parameters for M51 \label{paramtab}}
\tablehead{
\colhead{Parameter} & \colhead{Value} & \colhead{Reference}}
\startdata
Center RA ~($\alpha$) (J2000) & $13^h29^m52^s.71$  & \citet{Hagiwara01} \\ 
Center DEC ~($\delta$) (J2000) & $47^\circ11'42\farcs80$ & '' \\ 
Systematic Velocity ~($V_{sys}$)  & 472 (LSR) & \citet{Tully74sp} \\ 
Position Angle of Major Axis ~($\theta_{MA}$) & 170\degr & '' \\
Inclination ~($i$) & 20\degr & '' \\ 
\enddata
\end{deluxetable}

\begin{deluxetable}{lcc}
  \tablecolumns{3} \tablewidth{0pt} \tablecaption{Estimation of the
    Systematic Velocity of M51 \label{vsystab}} \tablehead{
    \colhead{Method} & \colhead{$V_{sys}$ (\kms)\tablenotemark{a}} &
    \colhead{Error (\kms)}} \startdata
  \citet{Tully74sp} & 472 & 3 \\
  Freeing $V_{sys}$ in fitting  &  470.6 & 0.4 \\
  for \vrad$(\psi)$ and \vtan$(\psi)$ &  &  \\
  Tilted Rings Analysis & 471.4 & 0.5 \\
  Rotation Curve Fitting & 473.3 & 0.5 \\
  \tableline
  Weighted Mean & 471.7 & 0.3 \\
  \enddata \tablenotetext{a}{Velocity in LSR frame}
\end{deluxetable}

\begin{deluxetable}{cccccc}
  \tablewidth{0pt} \tablecaption{Vortensity in the 47\arcsec\ -
    80\arcsec\ Annulus} \tablehead{ \colhead{Region} & \colhead
    {$\overline \Sigma$\tablenotemark{a}} & \colhead {$\left(
        \frac{v_\theta}{R} \right)$\tablenotemark{b}} & \colhead
    {$\left(\frac{\partial v_\theta}{\partial
          R}\right)$\tablenotemark{b}} &
    \colhead{$\left(\frac{1}{R}\frac{\partial v_R}{\partial
          \theta}\right)$\tablenotemark{b}} & \colhead{Vortensity
      Value\tablenotemark{c}}} \startdata

  Arm 1 ($\psi \approx 360\degr$) & 244 & 70 $\pm$ 1 & 103 $\pm$ 2 &
  -30
  $\pm$ 3 & 0.8 $\pm$ 0.2 \\
  Interarm 1 ($\psi \approx 90\degr$) & 16 & 67 $\pm$ 1 & -52 $\pm$ 3
  & 6 $\pm$ 1 & 0.5 $\pm$ 0.2
  \\
  Arm 2 ($\psi \approx 190\degr$)& 128 & 63 $\pm$ 0.5 & 59 $\pm$ 3 &
  -18
  $\pm$ 3 & 1 $\pm$ 0.2 \\
  Interarm 2 ($\psi \approx 275\degr$) & 19 & 59 $\pm$ 1 & -31 $\pm$ 2
  &
  11 $\pm$ 1 & 0.9 $\pm$ 0.2 \\
  \enddata {\singlespace \tablenotetext{a}{$\rm \left[M_\odot \,\,
        pc^{-2}\right]$; error of $\sim$20\%}
    \tablenotetext{b}{$\left[\rm km \,\, s^{-1} \,\,kpc^{-1}\right]$}
    \tablenotetext{c}{$\rm \left[ km \,\, s^{-1} \,\,kpc^{-1} \,
        (M_\odot\,\, pc^{-2})^{-1}\right]$} }
\label{vorttab}
\end{deluxetable}

\begin{deluxetable}{cccccc}
\tablewidth{0pt}
\tablecaption{Mass Flux in the 47\arcsec\ - 80\arcsec\ Annulus}
\tablehead{
\colhead{Region} & \colhead {$\Sigma$\tablenotemark{a}} &
\colhead {$ v_R \cos i_p$\tablenotemark{b}} & 
\colhead {$(v_\theta-\Omega_pR) \sin i_p$\tablenotemark{b}} &
\colhead {$C$\tablenotemark{c}} &
\colhead{$\Sigma\tilde{v_x}$\tablenotemark{d}}} 
\startdata
$\theta_{MA} = 170$\degr: \\
Arm 1 ($\psi \approx 360\degr$) & 244 &  -24 & 41 & -58 $\pm$ 30 & -10113 \\
Arm 2 ($\psi \approx 190\degr$)& 128 & -53 & 32 & -58 $\pm$ 30 & -10113  \\
Interarm 1 ($\psi \approx 90\degr$) & 16 & 23 & 18 & - & -259  \\ 
Interarm 2 ($\psi \approx 275\degr$) & 19 & 49 & 21 & - & 237 \\ 
\tableline
$\theta_{MA} = 175$\degr: \\
Arm 1 ($\psi \approx 360\degr$) & 244 & -41 & 41 & -32 $\pm$ 18 & -8103  \\
Arm 2 ($\psi \approx 190\degr$)& 128 & -63 & 32 & -32 $\pm$ 18 & -8103  \\
Interarm 1 ($\psi \approx 90\degr$) & 16 & 12 & 18 & - & -38  \\ 
Interarm 2 ($\psi \approx 275\degr$) & 19 & 36 & 21 & - & 464 \\ 
\tableline
$\theta_{MA} = 180$\degr: \\
Arm 1 ($\psi \approx 360\degr$) & 244 &  -59 & 41 & -8 $\pm$ 17 & -6345 \\
Arm 2 ($\psi \approx 190\degr$)& 128 & -71 & 32 & -8 $\pm$ 17 & -6345 \\
Interarm 1 ($\psi \approx 90\degr$) & 16 & 1 & 18 & - & 182 \\ 
Interarm 2 ($\psi \approx 275\degr$) & 19 & 27 & 21 & - & 761  \\ 
\enddata
{\singlespace
\tablenotetext{a}{$\left[\rm M_\odot \,\,pc^{-2}\right]$; error of $\sim$20\%}
\tablenotetext{b}{$\left[\rm km \,\, s^{-1} \right]$}
\tablenotetext{c}{$\left[\rm km \,\, s^{-1} \right]$; error largely
  due to errors in $\Sigma$ and $\Omega_p$}
\tablenotetext{d}{$\left[\rm M_\odot \,\, pc^{-2} \,\, km \,\, s^{-1} \right]$}
}
\label{mctab}
\end{deluxetable}

\clearpage

\begin{figure}
\includegraphics*[scale=0.80]{f17.eps}
\caption{\footnotesize H$\alpha$ \vrad\ fits as a function of arm
  phase for 3 different \pa s $\theta_{MA}$, 170\degr, 175\degr, and
  180\degr, for two annuli (47\arcsec\ - 80 \arcsec\ and 61\arcsec\ -
  105\arcsec).  We fix the inclination at 24\degr, and other
  parameters used in the fitting are shown in Table \ref{paramtab}.
  The dashed line is the mean CO intensity along the arm for a
  $\theta_{MA}$ of 170\degr, which varies only slightly with
  $\theta_{MA}$.  The error bars are not shown because they are
  similar to those shown in Figure \ref{Havans}.}
\label{pavr}
\end{figure}

\begin{figure}
\includegraphics*[angle=-90, scale=0.70]{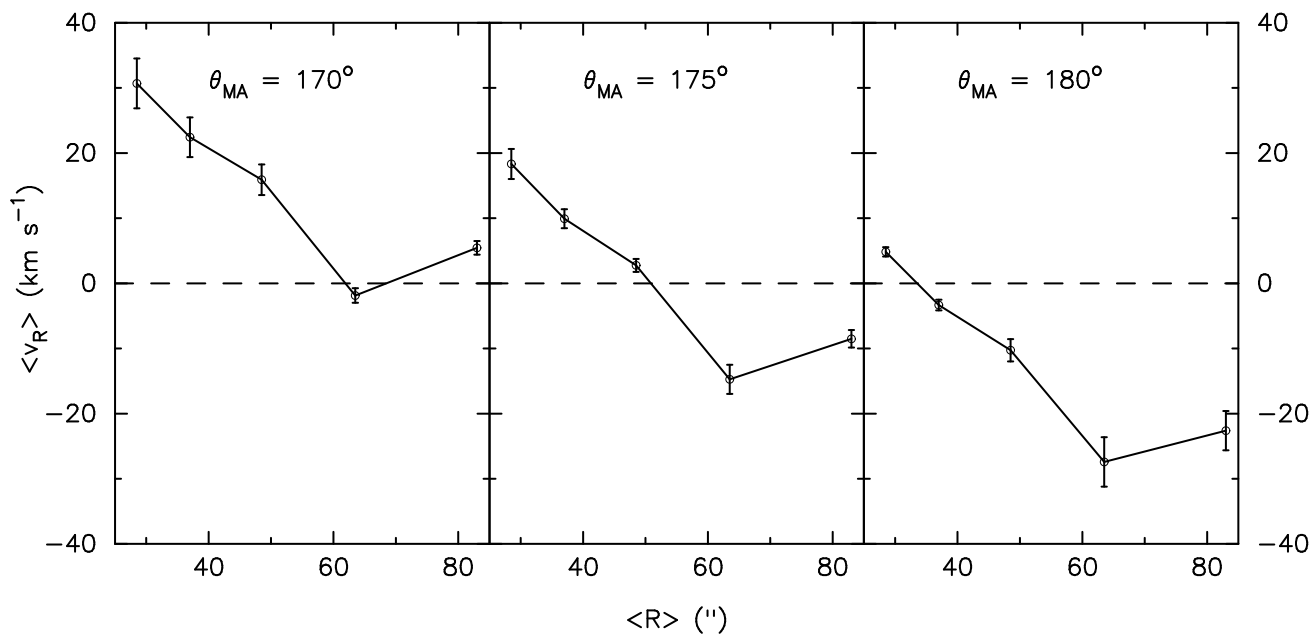}
\caption{\footnotesize Mass-weighted average radial velocities
  $\langle v_R \rangle$ in the different annuli, two of which are
  shown in Figure \ref{pavr}. The abscissa indicates the mean radius
  $\langle R \rangle$, in arcsecs, of each annulus.  The three panels
  show the mass-weighted average \vrad\ assuming three different
  values for the \pa\ $\theta_{MA}$.  The error bars include both
  fitted errors in \vrad\ (see Fig. \ref{COvans}) and an estimated
  error of 20\% in $\Sigma$.}
\label{flxvr}
\end{figure}

\end{document}